# On the optimal feedforward torque control problem of anisotropic synchronous machines: Quadrics, quartics and analytical solutions

Hisham Eldeeb[‡], Christoph M. Hackl[‡,⋆], Lorenz Horlbeck[†] and Julian Kullick[‡]

**Abstract**

The optimal feedforward torque control problem is tackled and solved analytically for synchronous machines while stator resistance and cross-coupling inductance are explicitly considered. Analytical solutions for the direct and quadrature optimal reference currents are found for all major operation strategies such as Maximum Torque per Current (MTPC) (or often called as Maximum Torque per Ampere (MTPA)), Maximum Current (MC), Field Weakening (FW), Maximum Torque per Voltage (MTPV) and Maximum Torque per Flux (MTPF). Numerical methods (approximating the optimal solutions only) or simplifying assumptions (neglecting stator resistance and/or cross-coupling inductance) are *no longer* necessary. The presented results are based on one simple idea: all optimization problems (e.g. MTPC, MTPV or MTPF) with their respective constraints (e.g. current or voltage limit) and the computation of the intersection point(s) of voltage ellipse, current circle, or torque, MTPC, MTPV and MTPF hyperbolas are reformulated implicitly as *quadrics* (quadratic surfaces) which allow to solve the feedforward torque control problem by invoking the Lagrangian formalism and by finding the roots of a fourth-order polynomial analytically. The proposed solutions are applicable to any anisotropic (or isotropic) synchronous machine independent of the underlying current control strategy.

## Notation

$\mathbb{N}, \mathbb{R}, \mathbb{C}$: natural, real, complex numbers. $\boldsymbol{x} := (x_1, \ldots, x_n)^\top \in \mathbb{R}^n$: column vector, $n \in \mathbb{N}$ where "$^\top$" and ":=" mean "transposed" (interchanging rows and columns of a matrix or vector) and "is defined as", resp., $\boldsymbol{0}_n \in \mathbb{R}^n$: zero vector. $\boldsymbol{a}^\top \boldsymbol{b} := a_1 b_1 + \cdots + a_n b_n$: scalar product of the vectors $\boldsymbol{a} := (a_1, \ldots, a_n)^\top$ and $\boldsymbol{b} := (b_1, \ldots, b_n)^\top$. $\|\boldsymbol{x}\| := \sqrt{\boldsymbol{x}^\top \boldsymbol{x}} = \sqrt{x_1^2 + \cdots + x_n^2}$: Euclidean norm of $\boldsymbol{x}$. $\boldsymbol{A} \in \mathbb{R}^{n \times n}$: (square) matrix with $n$ rows and columns. $\boldsymbol{A}^{-1}$: inverse of $\boldsymbol{A}$ (if exists). $\det(\boldsymbol{A})$: determinant of $\boldsymbol{A}$, $\mathrm{spec}(\boldsymbol{A})$: spectrum of $\boldsymbol{A}$ (the set of the eigenvalues of $\boldsymbol{A}$). $\boldsymbol{I}_n \in \mathbb{R}^{n \times n} := \mathrm{diag}(1, \ldots, 1)$: identity matrix. $\boldsymbol{O}_{n \times p} \in \mathbb{R}^{n \times p}$: zero matrix, $n, p \in \mathbb{N}$. $\boldsymbol{T}_\mathrm{p}(\phi_\mathrm{k}) = \left[\begin{smallmatrix} \cos(\phi_\mathrm{k}) & -\sin(\phi_\mathrm{k}) \\ \sin(\phi_\mathrm{k}) & \cos(\phi_\mathrm{k}) \end{smallmatrix}\right]$: Park transformation matrix (with electrical angle $\phi_\mathrm{k}$) and $\boldsymbol{J} := \boldsymbol{T}_\mathrm{p}(\pi/2) = \left[\begin{smallmatrix} 0 & -1 \\ 1 & 0 \end{smallmatrix}\right]$: rotation matrix (counter-clock wise rotation by $\frac{\pi}{2}$; see e.g. [1], [2]). "s.t.": subject to (optimization with constraints). $\jmath$: imaginary unit with $\jmath^2 = \sqrt{-1}$, "$\stackrel{!}{=}$": must equal. $X \cap Y$: intersection of the sets $X$ and $Y$ (in this paper: $X, Y \subset \mathbb{R}^2$).

## Contents

[‡]H. Eldeeb, C.M. Hackl and J. Kullick are with the research group "Control of renewable energy systems" (CRES) at the Munich School of Engineering (MSE), Technische Universität München (TUM), Germany.

[†]L. Horlbeck is with the Institute of Automotive Technology (FTM), Technische Universität München (TUM), Germany.

[*]Authors are in alphabetical order and contributed equally to the paper. Corresponding author is C.M. Hackl (christoph.hackl@tum.de).

## I. Introduction

The efficient use of electrical energy becomes more and more significant; in particular for electrical machines. Electrical machines consume more than half of the globally generated electricity [3]. Hence, advances in research on the control and modeling of electrical drives (machine+inverter) have been made; in particular for synchronous machines (SMs) with non-negligible anisotropy such as *interior permanent-magnet (PM) synchronous motors (IPMSMs)*, *reluctance synchronous machines (RSMs)*, *PM-assisted RSMs (PMA-RSMs)* or *PM-enhanced RSMs (PME-RSMs)* [4]–[6]. The two approaches for enhancing the efficiency of SMs are (a) improving the stator or rotor design of the electrical machine [4], [5], [7] or (b) extracting the highest possible efficiency by implementing an optimal feedforward torque control strategy (which is the topic under study in this paper).

Optimization of torque production while minimizing copper losses at *steady state* can be classified into two categories (i) "*search control*" (SC) and (ii) "*loss model control*" (LMC). The SC approach is considered as a *perturb and observe* adaptive strategy which continuously approximates the minimizing reference currents online for any given reference torque. The SC method does not require precise knowledge of the machine parameters [8]. Nevertheless, stability of such a strategy is not always guaranteed [9]. On the other hand, LMC–*adopted in this paper*–is based on the development of a mathematical model, that describes the electromechanical conversion and electrical losses of the machine during operation. Clearly, LMC strictly depends on the machine parameters which are sometimes provided by the manufacturer or can be obtained through experiments [10], [11]. Depending on the current and voltage constraints, and the actual angular velocity and the demanded reference torque of the electrical drive system, *four* optimal feedforward torque control strategies can be defined: (a) *Maximum Torque per Current (MTPC)* (which, in literature, is also known as *Maximum Torque per Ampere (MTPA)*), (b) *Maximum Current (MC)*, (c) *Field Weakening (FW)* and (d) *Maximum Torque per Voltage (MPTV)* or *Maximum Torque per Flux (MTPF)*.

The optimal feedforward torque control problem has been investigated in numerous publications, see e.g. [12]–[25] for MTPC, [7], [16], [17], [25]–[27] for FW, and [16], [27]–[30] for MTPV to name a few. Note that the MPTF operation strategy, which minimizes the stator iron losses at high speeds [15], [31], is a special (simplified) case of the MTPV operation strategy (shown later in this paper). The online computation of the optimal reference currents for the different operation strategies (such as MTPC, MC, FW, MTPV or MTPF) is usually done *numerically* or *analytically but with simplifying assumptions* on the machine model (e.g. neglecting stator resistance and/or cross-coupling inductance) or the physical constraints (e.g. voltage ellipse). Moreover, numerical solutions in general come at the expense of tightening the programmed tolerances which may decrease the speed of the control algorithm and increase the computational load on the real-time system. Analytical solutions are more attractive, since they are easier to implement, more accurate and faster to compute. However, to the best knowledge of the authors, analytical solutions *including* stator resistance and mutual inductance for MTPC, MC, FW, MTPV and MTPF of anisotropic SMs are scarcely investigated or not available at all. For example, in [15, Sec. IV], [28, Sec. 2.2.3], and [31, Sec. II-B], it was stated that acquiring a general analytical solution of the optimal currents for MTPV and MTPF while considering non-negligible stator resistance and magnetic cross-coupling is or seems *not possible* and would introduce a high degree of complexity.

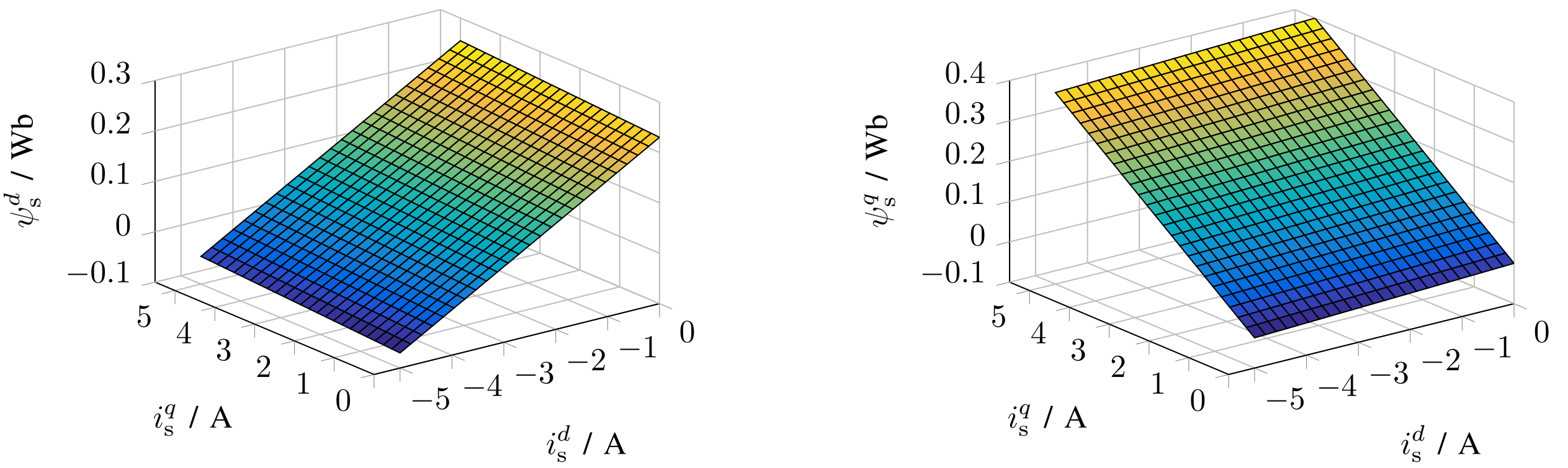

(a) *d-component* $\psi_{\rm s}^d$ *of the stator flux linkage.*

(b) *q-component* $\psi_{\rm s}^q$ *of the stator flux linkage.*

Figure 1: *Flux linkage map* (2) *of an exemplary SM with parameters as in* (36)*: Due to the non-zero mutual inductance, both flux linkage maps are slightly tilted (the map is only shown for the second quadrant, i.e.* $i_{\rm s}^d \leq 0$ *and* $i_{\rm s}^q \geq 0$*).*

Motivated by the aforementioned challenges and remaining open questions (e.g. how to consider stator resistance and cross-coupling in the whole feedforward torque control problem), the following research work has been conducted. The main contributions of this paper are:

(i) The derivation of a *unified theory for optimal feedforward torque control for anisotropic (and isotropic) SMs* which allows to compute the optimal reference currents *analytically* for *all* operation strategies (such as MTPC, MC, FW, MTPV and MTPF) incorporating *stator resistance* and *cross-coupling (mutual) inductance*. Moreover, the proposed analytical solutions offer (a) guaranteed convergence to the optimal reference currents (compared to numerical methods), (b) easy and straight-forward implementation, and (c) rapid execution and low computational burden making the proposed analytical algorithms very suitable even for modest (hence cheap) processor boards.

(ii) The unified theory is established by (a) the use of *Lagrangian multipliers* and (b) an *implicit* problem formulation as *quadrics* (i.e. all trajectories of constraints and operation strategies in the $(d, q)$-plane–such as e.g. current circle, voltage ellipse and torque hyperbola–are reformulated implicitly as quadric surfaces).

(iii) It is shown that *only* for very high speeds or very small stator resistances, the MTPF solution is an acceptable approximation of the MTPV solution. In general, an MTPV algorithm incorporating stator resistance and mutual inductance yields higher efficiencies and should be preferred and implemented.

(iv) The negative effects of neglecting stator resistance and mutual inductance or both on the optimality of all operation strategies are illustrated, which show that neglecting these two parameters during optimization will lead to significant deviations between optimal and approximated reference currents.

The remainder of the paper is organized as follows: Section II re-visits the dynamic model of anisotropic synchronous machines, its operation constraints (such as current and voltage constraint) and the problem formulation of optimal feedforward torque control. Section III deals with mathematical preliminaries at steady state and the implicit reformulations (as quadrics) of the optimization problems and machine constraints. This paves the way for Section IV, where the *analytical* solutions for MTPC, MC, FW, MTPV and MTPF are presented. To improve readability, the thorough mathematical derivations of the unified theory are collected and summarized in the Appendices A-D.

## II. Problem formulation

### A. Generic dynamical model of synchronous machines (SMs)

The model of an anisotropic (a) permanent-magnet synchronous machine (PMSM), (b) permanent-magnet assisted or excited reluctance synchronous machine (PMA-RSM or PME-RSM, resp. [6], [23]) or (c) reluctance synchronous machine (RSM [6], [32], [33]) in the $(d, q)$-reference frame[1] is given by

$$\left.\begin{aligned}
&\overbrace{\begin{pmatrix} u_{\rm s}^d(t) \\ u_{\rm s}^q(t)\end{pmatrix}}^{=:\boldsymbol{u}_{\rm s}^k(t)} = R_{\rm s} \overbrace{\begin{pmatrix} i_{\rm s}^d(t) \\ i_{\rm s}^q(t)\end{pmatrix}}^{=:\boldsymbol{i}_{\rm s}^k(t)} + \omega_{\rm k}(t) \overbrace{\begin{bmatrix} 0 & -1 \\ 1 & 0 \end{bmatrix}}^{=:\boldsymbol{J}} \overbrace{\begin{pmatrix} \psi_{\rm s}^d\big(\boldsymbol{i}_{\rm s}^k(t)\big) \\ \psi_{\rm s}^q\big(\boldsymbol{i}_{\rm s}^k(t)\big)\end{pmatrix}}^{=:\boldsymbol{\psi}_{\rm s}^k\left(\boldsymbol{i}_{\rm s}^k(t)\right)} + \tfrac{\rm d}{{\rm d}t}\boldsymbol{\psi}_{\rm s}^k(\boldsymbol{i}_{\rm s}^k(t)), \qquad && \boldsymbol{\psi}_{\rm s}^k(\boldsymbol{i}_{\rm s}^k(0)) = \boldsymbol{\psi}_{\rm s}^{k,0} \in \mathbb{R}^2 \\
&\tfrac{\rm d}{{\rm d}t}\omega_{\rm k}(t) = \frac{n_{\rm p}}{\Theta}\Big(m_{\rm m}(\boldsymbol{i}_{\rm s}^k(t)) - m_{\rm l}(t)\Big), && \omega_{\rm k}(0) = n_{\rm p}\omega_{\rm m}^0 \in \mathbb{R} \\
&\tfrac{\rm d}{{\rm d}t}\phi_{\rm k}(t) = \omega_{\rm k}(t), && \phi_{\rm k}(0) = n_{\rm p}\phi_{\rm m}^0 \in \mathbb{R}.
\end{aligned}\right\} \tag{1}$$

[1] I.e., the synchronously rotating $k = (d, q)$-coordinate system with orthogonal axes $d$ and $q$ after Clarke and Park transformation [1], [2].

with initial values $\psi_s^{k,0}$, $\omega_m^0$ and $\phi_m^0$. The following assumption is imposed on the flux linkage:

**Assumption II.1.** *For a (locally)* constant *inductance matrix* $\boldsymbol{L}_s^k \in \mathbb{R}^{2\times 2}$ *(e.g. obtained by a linearization of* (1) *at the actual operating point), the (local approximation of the) flux linkage can be expressed as follows*

$$\boldsymbol{\psi}_s^k(\boldsymbol{i}_s^k(t)) = \underbrace{\begin{bmatrix} L_s^d & L_m \\ L_m & L_s^q \end{bmatrix}}_{=:\boldsymbol{L}_s^k \in \mathbb{R}^{2\times 2}} \boldsymbol{i}_s^k(t) + \underbrace{\begin{pmatrix} \psi_{pm}^d \\ \psi_{pm}^q \end{pmatrix}}_{=:\boldsymbol{\psi}_{pm}^k} \quad \textit{where} \quad \boldsymbol{\psi}_{pm}^k = \begin{cases} (\psi_{pm}, 0)^\top, & \textit{for PMSM and PME-RSM [6]} \\ (0, -\psi_{pm})^\top, & \textit{for PMA-RSM [6], [23]} \\ (0,0)^\top, & \textit{for RSM [6], [32], [33].} \end{cases} \tag{2}$$

Then, the machine torque can be computed as follows

$$m_m(\boldsymbol{i}_s^k(t)) = \tfrac{3}{2} n_p \, \boldsymbol{i}_s^k(t)^\top \boldsymbol{J} \boldsymbol{\psi}_s^k(\boldsymbol{i}_s^k(t)) \overset{(2)}{=} \tfrac{3}{2} n_p \left[ \boldsymbol{i}_s^k(t)^\top \boldsymbol{J} \boldsymbol{L}_s^k \boldsymbol{i}_s^k(t) + \boldsymbol{i}_s^k(t)^\top \boldsymbol{J} \boldsymbol{\psi}_{pm}^k \right] \tag{3}$$

$$= \tfrac{3}{2}\, n_p \left[ \psi_{pm}^d i_s^q(t) - \psi_{pm}^q i_s^d(t) + \left(L_s^d - L_s^q\right) i_s^d(t) i_s^q(t) + L_m \left( i_s^q(t)^2 - i_s^d(t)^2 \right) \right] \tag{4}$$

$$\overset{(2)}{=} \begin{cases} \tfrac{3}{2}\, n_p \left[ \psi_{pm} i_s^q(t) + \left(L_s^d - L_s^q\right) i_s^d(t) i_s^q(t) + L_m \left( i_s^q(t)^2 - i_s^d(t)^2 \right) \right], & \text{for PMSM or PME-RSM} \\ \tfrac{3}{2}\, n_p \left[ \psi_{pm} i_s^d(t) + \left(L_s^d - L_s^q\right) i_s^d(t) i_s^q(t) + L_m \left( i_s^q(t)^2 - i_s^d(t)^2 \right) \right], & \text{for PMA-RSM,} \\ \tfrac{3}{2}\, n_p \left[ \left(L_s^d - L_s^q\right) i_s^d(t) i_s^q(t) + L_m \left( i_s^q(t)^2 - i_s^d(t)^2 \right) \right], & \text{for RSM.} \end{cases}$$

In (1), (2) and (3) or (4), $R_s$ (in $\Omega$) is the stator resistance, $\boldsymbol{u}_s^k := (u_s^d, u_s^q)^\top$ (in V), $\boldsymbol{i}_s^k := (i_s^d, i_s^q)^\top$ (in A) and $\boldsymbol{\psi}_s^k := (\psi_s^d, \psi_s^q)^\top$ (in Wb) are stator voltage (e.g. applied by a voltage source inverter), current and flux linkage vectors in the $(d,q)$-reference frame, respectively. Note that $\omega_k = n_p \omega_m$ (in rad/s) and $\phi_k = n_p \phi_m$ are *electrical* angular frequency and angle, whereas $\omega_m$ and $\phi_m$ are *mechanical* angular frequency and angle of the rotor (with initial values $\omega_m^0$ and $\phi_m^0$), respectively. $n_p$ is the pole pair number of the machine and $\Theta$ (in $\mathrm{kg\,m^2}$) is the (rotor's) inertia. $m_m$ is the electro-magnetic machine torque[2] and $m_l$ (in $\mathrm{Nm}$) is a (bounded) load torque. The flux linkage $\boldsymbol{\psi}_s^k$ depends on the *symmetric*, *positive-definite* inductance matrix $\boldsymbol{L}_s^k = (\boldsymbol{L}_s^k)^\top > 0$ [32] with stator inductances $L_s^q > 0$, $L_s^d > 0$ (both in H) and cross-coupling (mutual) inductance[3] $L_m \in \mathbb{R}$ (in H) and $L_s^d L_s^q - L_m^2 > 0$, the stator currents $\boldsymbol{i}_s^k$ and the permanent-magnet flux linkage $\boldsymbol{\psi}_{pm}^k = (\psi_{pm}^d, \psi_{pm}^q)^\top$. Fig. 1 illustrates the flux linkage (2) of an anisotropic IPMSM for some exemplified values of $L_s^q$, $L_s^d$, $L_m$ and $\psi_{pm}$.

**Remark II.2** (Affine flux linkage)**.** *Note that Assumption II.1 implies a (locally) constant inductance matrix; this is in line with most recent publications (even from 2016) which also deal with constant inductances only (see e.g. [12], [17], [24], [27], [35]–[37] for PMSMs or [28], [38] for RSMs). This simplification will* not *be true in the most general case when the flux linkage is a nonlinear function of the currents [11], [33]. Nevertheless, in the humble opinion of the authors, the presented results are of quite some relevance and have not been discussed in this general framework before: The results of this paper can be considered as a generalization of the results for IPMSM in [16] by including resistance $R_s$ and mutual inductance $L_m$ into the generic optimization formulation. The simplifying assumptions which neglect these physical parameters are overcome. The consideration of nonlinear flux linkages is ongoing research.*

**Remark II.3** (Inductance ratios and signs of permanent-magnet flux linkage)**.** *For different machine designs, the direct and quadrature inductances take different values and have different ratios; also the permanent-magnet flux constant changes its sign [6], i.e.*

- *PMSMs:* $\psi_{pm}^d > 0$, $\psi_{pm}^q = 0$ *and* $L_s^q \geq L_s^q \Leftrightarrow \frac{L_s^d}{L_s^q} \leq 1$ *(inverse saliency ratio);*
- *PME-RSMs:* $\psi_{pm}^d > 0$, $\psi_{pm}^q = 0$ *and (a)* $L_s^q \geq L_s^q \Leftrightarrow \frac{L_s^d}{L_s^q} \leq 1$ *(inverse saliency ratio) or (b)* $L_s^q \leq L_s^q \Leftrightarrow \frac{L_s^d}{L_s^q} \geq 1$ *(normal saliency ratio);*
- *PMA-RSMs with normal saliency:* $\psi_{pm}^d = 0$, $\psi_{pm}^q < 0$ *and* $L_s^q \leq L_s^q \Leftrightarrow \frac{L_s^d}{L_s^q} \geq 1$ *(normal saliency ratio); and*
- *RSMs:* $\psi_{pm}^d = \psi_{pm}^q = 0$ *and* $L_s^q \leq L_s^q \Leftrightarrow \frac{L_s^d}{L_s^q} \geq 1$ *(normal saliency ratio);*

### *B. Operation constraints*

Due to safety and physical reasons [34, Cha. 16], stator currents and voltages should never exceed their respective maximal amplitudes $\hat{\imath}_{max} > 0$ (in A) and $\hat{u}_{max} > 0$ (in V). Hence, the following must be ensured by the control system for all time

$$\forall t \geq 0\,\mathrm{s}: \qquad \|\boldsymbol{i}_s^k(t)\|^2 = i_s^d(t)^2 + i_s^q(t)^2 \leq \hat{\imath}_{max}(t)^2 \qquad \text{and} \qquad \|\boldsymbol{u}_s^k(t)\|^2 = u_s^d(t)^2 + u_s^q(t)^2 \leq \hat{u}_{max}(t)^2. \tag{5}$$

[2] The factor 3/2 is due to an amplitude-correct Clarke transformation [34, Sec. 16.7].

[3] Note that the mutual inductance $L_m$ changes its sign with the negative product of the currents, i.e. $\mathrm{sign}(L_m) = -\,\mathrm{sign}(i_s^d \cdot i_s^q)$ [33, Fig. 2].

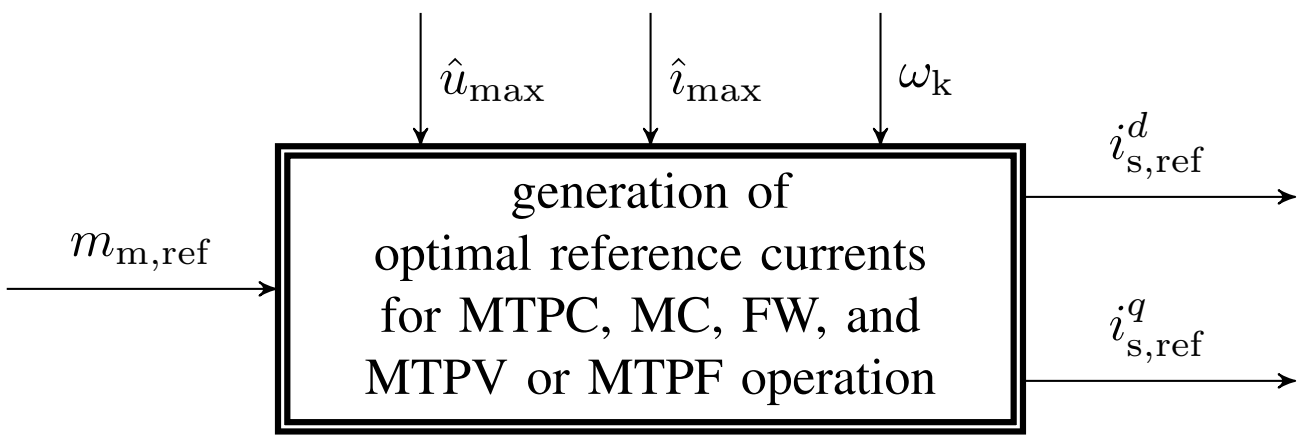

Figure 2: *Optimal reference current generation for a given reference torque $m_{\rm m,ref}$: The computed, optimal reference currents $i^d_{\rm s,ref}$ and $i^q_{\rm s,ref}$ depend also on maximal current $\hat{\imath}_{\max}$, maximal voltage $\hat{u}_{\max}$ and (actual) electrical angular frequency $\omega_{\rm k} = n_{\rm p}\omega_{\rm m}$.*

Note that maximally admissible current $\hat{\imath}_{\max}$ and voltage $\hat{u}_{\max}$ might change over time: (a) the current limit is usually equal to the nominal/rated current of the machine but can also exceed this nominal value for a short period in time and (b) the maximally applicable voltage will change with the DC-link voltage of the inverter. In the remainder of this paper, the time-dependency will not be highlighted explicitly and the argument $(t)$ will be dropped.

### C. Problem formulation

For a given reference torque $m_{\rm m,ref}$ (in $\rm N\,m$) (and given current and voltage limits), the general control objective is to find analytical solutions of the optimal reference currents for all operation modes (such as e.g. MTPC, MTPF and MTPV). From a mathematical point of view, the following optimization problem

$$\max_{\boldsymbol{i}^k_{\rm s}} \; -f(\boldsymbol{i}^k_{\rm s}) \;\; \text{s.t.} \;\; \|\boldsymbol{u}^k_{\rm s}\| \leq \hat{u}_{\max}, \; \|\boldsymbol{i}^k_{\rm s}\| \leq \hat{\imath}_{\max}, \; |m_{\rm m}(\boldsymbol{i}^k_{\rm s})| \leq |m_{\rm m,ref}| \;\; \text{and} \;\; \text{sign}(m_{\rm m,ref}) = \text{sign}(m_{\rm m}(\boldsymbol{i}^k_{\rm s})) \tag{6}$$

with three inequality constraints and one equality constraint must be solved online, where obviously the sign of reference and machine torque should coincide. The function $f(\boldsymbol{i}^k_{\rm s})$ depends on the operation strategy (e.g. $f(\boldsymbol{i}^k_{\rm s}) = \|\boldsymbol{i}^k_{\rm s}\|^2$ for MTPC; for more details, see Sec. IV), The most favorable outcome is an analytical solution which gives explicitly the reference current vector

$$\boxed{\boldsymbol{i}^k_{\rm s,ref}(m_{\rm m,ref}, \hat{u}_{\max}, \hat{\imath}_{\max}, \omega_{\rm k}) = \begin{pmatrix} i^d_{\rm s,ref}(m_{\rm m,ref}, \hat{u}_{\max}, \hat{\imath}_{\max}, \omega_{\rm k}) \\ i^q_{\rm s,ref}(m_{\rm m,ref}, \hat{u}_{\max}, \hat{\imath}_{\max}, \omega_{\rm k}) \end{pmatrix} := \arg\max_{\boldsymbol{i}^k_{\rm s}} m_{\rm m}(\boldsymbol{i}^k_{\rm s})} \tag{7}$$

as function (see Fig. 2) of reference torque $m_{\rm m,ref}$ (coming from an outer control loop; e.g. the speed control loop), voltage limit $\hat{u}_{\max}$, current limit $\hat{\imath}_{\max}$ and electrical angular velocity $\omega_{\rm k} = n_{\rm p}\omega_{\rm m}$. The computed reference current vector $\boldsymbol{i}^k_{\rm s,ref}$ can then directly be handed over to any underlying current control loop(s).

**Remark II.4** (Feasible reference torques and non-convexity of the machine torque)**.** *Note that, due to the voltage limit (during high-speed operation) or due to the current limit, the range of admissible reference torques is restricted. Hence, not all reference torques $m_{\rm m,ref}$ are feasible during all operation modes, therefore the additional inequality constraint in* (6) *must be added. If the requested reference torque is feasible, the inequality constraint becomes the equality constraint $|m_{\rm m}(\boldsymbol{i}^k_{\rm s})| = |m_{\rm m,ref}|$ (or simply, $m_{\rm m}(\boldsymbol{i}^k_{\rm s}) = m_{\rm m,ref}$). Important to note that the machine torque $m_{\rm m}(\boldsymbol{i}^k_{\rm s}) \propto (\boldsymbol{i}^k_{\rm s})^\top \boldsymbol{J}\boldsymbol{L}^k_{\rm s}\boldsymbol{i}^k_{\rm s}$ is* not *convex, since $\boldsymbol{J}\boldsymbol{L}^k_{\rm s} = \left[\begin{smallmatrix} -L_{\rm m} & -L^q_{\rm s} \\ L^d_{\rm s} & L_{\rm m} \end{smallmatrix}\right]$ is non-symmetric and indefinite with eigenvalues $\pm\sqrt{L^2_{\rm m} - L^d_{\rm s}L^q_{\rm s}}$. Hence, maximizing the machine torque is not a viable approach. To account for that, the general optimization problem* (6) *will be divided into several sub-problems leading to the optimal operation strategies MTPC, MC, FW, MTPV or MTPF.*

## III. Mathematical preliminaries: Steady-state operation and implicit formulations

### A. Steady-state operation

In the remainder of this paper, only steady-state operation will be considered; which implies that $\frac{\rm d}{{\rm d}t}\boldsymbol{\psi}^k_{\rm s}(\boldsymbol{i}^k_{\rm s}) = \frac{\rm d}{{\rm d}t}\boldsymbol{i}^k_{\rm s} = \boldsymbol{0}_2$. Inserting (2) into (1) and neglecting the time derivative of the current, i.e. $\frac{\rm d}{{\rm d}t}\boldsymbol{i}^k_{\rm s} = \boldsymbol{0}_2$, the steady-state stator circuit model of a SM in matrix/vector notation is obtained as follows

$$\boldsymbol{u}^k_{\rm s} = R_{\rm s}\boldsymbol{i}^k_{\rm s} + \omega_{\rm k}\boldsymbol{J}\boldsymbol{L}^k_{\rm s}\boldsymbol{i}^k_{\rm s} + \omega_{\rm k}\boldsymbol{J}\boldsymbol{\psi}^k_{\rm pm} \tag{8}$$

where $\boldsymbol{J}$, and $\boldsymbol{L}^k_{\rm s}$ and $\boldsymbol{\psi}^k_{\rm pm}$ are as in (1) and (2), respectively.

### B. Implicit formulation of machine torque and constraints as quadrics

The steady-state SM model (8) will be the basis for all upcoming derivations. The trick to obtain and derive a unified theory for the optimal torque control problem under current and voltage constraints is the re-formulation of the general optimization problem (6) *implicitly by quadrics (or quadric surfaces)* which will allow to invoke the Lagrangian formalism to derive an analytical solution for all different operation modes (such as MTPC, MTPV, FW, etc.). In the upcoming sections, the implicit forms of torque hyperbola, voltage ellipse (elliptical area), current circle (circular area) and flux norm are presented. The explicit forms are also given (as link to the existing literature) if their expressions are not too long. Stator resistance $R_\mathrm{s} \neq 0$ and mutual inductance $L_\mathrm{m} \neq 0$ will *not* be neglected to present the most general result achievable within the framework of affine flux linkages as in (2).

*1) Torque hyperbola (constant torque trajectory):* To derive the implicit form as quadric of the torque hyperbola, the following symmetric matrix, vector and constant are defined:

$$\left.\begin{aligned} \boldsymbol{T} &:= \tfrac{3}{4} n_\mathrm{p}\big(\boldsymbol{J}\boldsymbol{L}_\mathrm{s}^k + \boldsymbol{L}_\mathrm{s}^k\boldsymbol{J}^\top\big) = \tfrac{3}{2} n_\mathrm{p} \begin{bmatrix} -L_\mathrm{m} & \frac{L_\mathrm{s}^d - L_\mathrm{s}^q}{2} \\ \frac{L_\mathrm{s}^d - L_\mathrm{s}^q}{2} & L_\mathrm{m} \end{bmatrix} = \boldsymbol{T}^\top, \\ \boldsymbol{t} &:= \tfrac{3}{4} n_\mathrm{p} \boldsymbol{J}\boldsymbol{\psi}_\mathrm{pm}^k = \tfrac{3}{2} n_\mathrm{p} \begin{pmatrix} -\frac{\psi_\mathrm{pm}^q}{2} \\ \frac{\psi_\mathrm{pm}^d}{2} \end{pmatrix} \overset{(2)}{=} \begin{cases} \tfrac{3}{2} n_\mathrm{p} \left(0,\ \frac{\psi_\mathrm{pm}}{2}\right)^\top, & \text{for PMSM and PME-RSM} \\ \tfrac{3}{2} n_\mathrm{p} \left(\frac{\psi_\mathrm{pm}}{2},\ 0\right)^\top, & \text{for PMA-RSM} \\ (0,\,0)^\top, & \text{for RSM,} \end{cases} \\ \tau(m_\mathrm{m,ref}) &:= -m_\mathrm{m,ref}. \end{aligned}\right\} \tag{9}$$

Moreover, note that $(\boldsymbol{i}_\mathrm{s}^k)^\top \boldsymbol{J}\boldsymbol{L}_\mathrm{s}^k \boldsymbol{i}_\mathrm{s}^k = (\boldsymbol{i}_\mathrm{s}^k)^\top \boldsymbol{L}_\mathrm{s}^k \boldsymbol{J}^\top \boldsymbol{i}_\mathrm{s}^k$ , hence

$$\tfrac{3}{4} n_\mathrm{p} (\boldsymbol{i}_\mathrm{s}^k)^\top (\boldsymbol{J}\boldsymbol{L}_\mathrm{s}^k + \boldsymbol{L}_\mathrm{s}^k \boldsymbol{J}^\top) \boldsymbol{i}_\mathrm{s}^k = (\boldsymbol{i}_\mathrm{s}^k)^\top \boldsymbol{T} \boldsymbol{i}_\mathrm{s}^k. \tag{10}$$

Now, by combining (9) and (10) with (3), the machine torque can be written as quadric as follows

$$m_\mathrm{m}(\boldsymbol{i}_\mathrm{s}^k) = (\boldsymbol{i}_\mathrm{s}^k)^\top \boldsymbol{T} \boldsymbol{i}_\mathrm{s}^k + 2\boldsymbol{t}^\top \boldsymbol{i}_\mathrm{s}^k. \tag{11}$$

For machine torque $m_\mathrm{m}(\boldsymbol{i}_\mathrm{s}^k)$ as in (11) and *constant* reference torque $m_\mathrm{m,ref}$, the machine torque trajectory can be expressed implicitly as *torque hyperbola* by invoking (9) as follows

$$\boxed{\mathbb{T}(m_\mathrm{m,ref}) := \left\{ \boldsymbol{i}_\mathrm{s}^k \in \mathbb{R}^2 \;\middle|\; (\boldsymbol{i}_\mathrm{s}^k)^\top \boldsymbol{T} \boldsymbol{i}_\mathrm{s}^k + 2\boldsymbol{t}^\top \boldsymbol{i}_\mathrm{s}^k + \tau(m_\mathrm{m,ref}) = 0 \right\}.} \tag{12}$$

Note that each summand in (12) has the unit VAs = Nm. An exemplary torque hyperbola is plotted in Fig. 3 (see —— line).

**Remark III.1** (Explicit expression for the torque hyperbola)**.** *For e.g. $i_\mathrm{s}^d \leq 0$ and $m_\mathrm{m,ref} > 0$, the torque hyperbola can be expressed explicitly (by solving* (4) *for $i_\mathrm{s}^q$):*

$$\mathbb{T}(i_\mathrm{s}^d, m_\mathrm{m,ref}) = \begin{cases} \dfrac{1}{3} \dfrac{3\psi_\mathrm{pm}^q n_\mathrm{p} i_\mathrm{s}^d + 2m_\mathrm{m,ref}}{n_\mathrm{p}\big[\psi_\mathrm{pm}^d + (L_\mathrm{s}^d - L_\mathrm{s}^q)\, i_\mathrm{s}^d\big]}, & L_\mathrm{m} = 0 \\[2ex] -\dfrac{(L_\mathrm{s}^d - L_\mathrm{s}^q) i_\mathrm{s}^d + \psi_\mathrm{pm}^d}{2L_\mathrm{m}} + \\ +\dfrac{\sqrt{[(L_\mathrm{s}^d - L_\mathrm{s}^q)^2 + 4L_\mathrm{m}^2]\,(i_\mathrm{s}^d)^2 + 2\psi_\mathrm{pm}^d (L_\mathrm{s}^d - L_\mathrm{s}^q)\, i_\mathrm{s}^d + 4L_\mathrm{m}(i_\mathrm{s}^d \psi_\mathrm{pm}^q + \frac{2m_\mathrm{m,ref}}{3p}) + (\psi_\mathrm{pm}^d)^2}}{2L_\mathrm{m}} & L_\mathrm{m} \neq 0, \end{cases} \tag{13}$$

*which holds for all $i_\mathrm{s}^d \neq \psi_\mathrm{pm}/(L_\mathrm{s}^d - L_\mathrm{s}^q)$. Note that the explicit expression* (13) *relies on a case study for the mutual inductance (and the signs of current $i_\mathrm{s}^d$ and reference torque $m_\mathrm{m,ref}$). The implicit form* (12) *holds in general and can easily be plotted (e.g. in Matlab using the command* `ezplot`*).*

*2) Voltage elliptical area (reformulation of the voltage constraint in* (5)*):* Recall that $\boldsymbol{J}^\top \boldsymbol{J} = \boldsymbol{I}_2$, $\alpha^\top = \alpha \in \mathbb{R}$ (the transpose of a scalar is the scalar itself), $(\boldsymbol{M}\boldsymbol{N})^\top = \boldsymbol{N}^\top \boldsymbol{M}^\top$ (for matrices of appropriate size) and $(\boldsymbol{L}_\mathrm{s}^k)^\top = \boldsymbol{L}_\mathrm{s}^k > 0$. With that in mind, inserting (8) into (5) and squaring the result yield

$$\begin{aligned} \hat{u}_\mathrm{max}^2 &\geq \|\boldsymbol{u}_\mathrm{s}^k\|^2 = (\boldsymbol{u}_\mathrm{s}^k)^\top \boldsymbol{u}_\mathrm{s}^k = (u_\mathrm{s}^d)^2 + (u_\mathrm{s}^q)^2 \\ &\overset{(8)}{\geq} R_\mathrm{s}^2 (\boldsymbol{i}_\mathrm{s}^k)^\top \boldsymbol{I}_2 \boldsymbol{i}_\mathrm{s}^k + R_\mathrm{s}\omega_\mathrm{k} (\boldsymbol{i}_\mathrm{s}^k)^\top \boldsymbol{J}\boldsymbol{L}_\mathrm{s}^k \boldsymbol{i}_\mathrm{s}^k + R_\mathrm{s}\omega_\mathrm{k} (\boldsymbol{i}_\mathrm{s}^k)^\top \boldsymbol{J}\boldsymbol{\psi}_\mathrm{pm}^k + \omega_\mathrm{k}^2 (\boldsymbol{i}_\mathrm{s}^k)^\top \underbrace{(\boldsymbol{L}_\mathrm{s}^k)^\top \boldsymbol{J}^\top \boldsymbol{J} \boldsymbol{L}_\mathrm{s}^k}_{=(\boldsymbol{L}_\mathrm{s}^k)^2} \boldsymbol{i}_\mathrm{s}^k + R_\mathrm{s}\omega_\mathrm{k} (\boldsymbol{i}_\mathrm{s}^k)^\top \boldsymbol{L}_\mathrm{s}^k \boldsymbol{J}^\top \boldsymbol{i}_\mathrm{s}^k \\ &\quad + \omega_\mathrm{k}^2 (\boldsymbol{i}_\mathrm{s}^k)^\top (\boldsymbol{L}_\mathrm{s}^k)^\top \boldsymbol{J}^\top \boldsymbol{J} \boldsymbol{\psi}_\mathrm{pm}^k + R_\mathrm{s}\omega_\mathrm{k} (\boldsymbol{\psi}_\mathrm{pm}^k)^\top \boldsymbol{J}^\top \boldsymbol{i}_\mathrm{s}^k + \omega_\mathrm{k}^2 (\boldsymbol{\psi}_\mathrm{pm}^k)^\top \boldsymbol{J}^\top \boldsymbol{J} \boldsymbol{L}_\mathrm{s}^k \boldsymbol{i}_\mathrm{s}^k + \omega_\mathrm{k}^2 \underbrace{(\boldsymbol{\psi}_\mathrm{pm}^k)^\top \boldsymbol{J}^\top \boldsymbol{J} \boldsymbol{\psi}_\mathrm{pm}^k}_{=\|\boldsymbol{\psi}_\mathrm{pm}^k\|^2 = \psi_\mathrm{pm}^2}. \end{aligned} \tag{14}$$

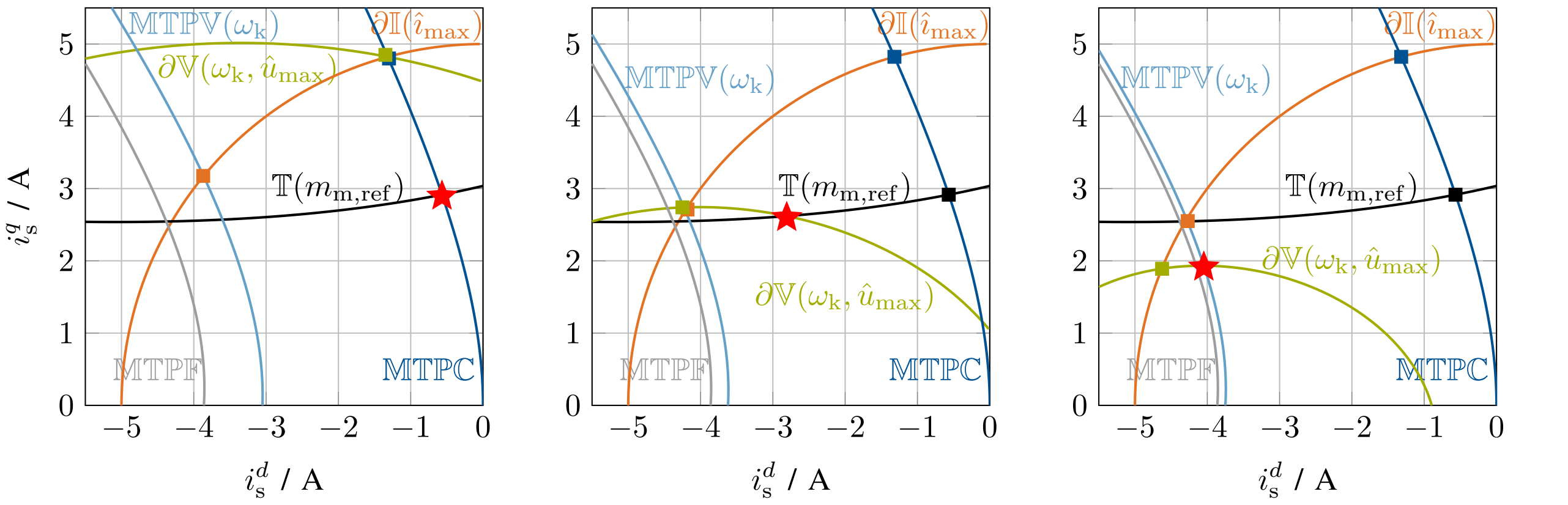

(a) *Maximum-Torque-per-Current (MTPC), i.e.,* $\mathbb{MTPC} \cap \mathbb{T}(m_{\mathrm{m,ref}})$*, for* $\omega_{\mathrm{k}} = \omega_{\mathrm{k,nom}}^{\mathrm{MC}} \leq \omega_{\mathrm{k,nom}}$.

(b) *Field weakening (FW), i.e.,* $\mathbb{T}(m_{\mathrm{m,ref}}) \cap \partial\mathbb{V}(\omega_{\mathrm{k}}, \hat{u}_{\max})$*, for* $\omega_{\mathrm{k,nom}} < \omega_{\mathrm{k}} = 2\omega_{\mathrm{k,nom}} < \omega_{\mathrm{k,cut\text{-}in}}^{\mathrm{MTPV}}$.

(c) *Maximum-Torque-per-Voltage (MTPV), i.e.,* $\mathbb{MTPV}(\omega_{\mathrm{k}}) \cap \partial\mathbb{V}(\omega_{\mathrm{k}}, \hat{u}_{\max})$*, for* $\omega_{\mathrm{k}} = 3\omega_{\mathrm{k,nom}} \geq \omega_{\mathrm{k,cut\text{-}in}}^{\mathrm{MTPV}}$ *(*$m_{\mathrm{m,ref}}$ *is* not *feasible anymore).*

Figure 3: *Illustration of three different feedforward torque control strategies for torque reference* $m_{\mathrm{m,ref}} > 0$ *(2nd quadrant) and varying electrical angular velocities (a)* $\omega_{\mathrm{k}} = \omega_{\mathrm{k,nom}}$*, (b)* $\omega_{\mathrm{k}} = 2\omega_{\mathrm{k,nom}}$ *and (c)* $\omega_{\mathrm{k}} = 3\omega_{\mathrm{k,nom}}$*: The three plots show voltage ellipse* — $\partial\mathbb{V}(\omega_{\mathrm{k}}, \hat{u}_{\max})$*, maximum current circle* — $\partial\mathbb{I}(\hat{\imath}_{\max})$*, MTPC hyperbola* — $\mathbb{MTPC}$*, torque hyperbola* — $\mathbb{T}(m_{\mathrm{m,ref}})$*, MTPV hyperbola* — $\mathbb{MTPV}(\omega_{\mathrm{k}})$*, MTPF hyperbola* — $\mathbb{MTPF}$ *and optimal operation point* ★ *, respectively.*

To find a more compact representation, the goal is to rewrite (14) as a quadric. Therefore, in (14), terms of the form $(\boldsymbol{i}_{\mathrm{s}}^k)^\top \square \boldsymbol{i}_{\mathrm{s}}^k$ and $\square^\top \boldsymbol{i}_{\mathrm{s}}^k$ (where $\square$ is either a matrix or a vector, respectively) are collected. Then, by defining

$$\left.\begin{aligned}
\boldsymbol{V}(\omega_{\mathrm{k}}) \;&:=\; \begin{bmatrix} v_{11}(\omega_{\mathrm{k}}), & v_{12}(\omega_{\mathrm{k}}) \\ v_{12}(\omega_{\mathrm{k}}), & v_{22}(\omega_{\mathrm{k}}) \end{bmatrix} = R_{\mathrm{s}}^2 \boldsymbol{I}_2 + R_{\mathrm{s}}\omega_{\mathrm{k}}\big(\boldsymbol{J}\boldsymbol{L}_{\mathrm{s}}^k + \boldsymbol{L}_{\mathrm{s}}^k\boldsymbol{J}^\top\big) + \omega_{\mathrm{k}}^2(\boldsymbol{L}_{\mathrm{s}}^k)^2 \\
&=\; \begin{bmatrix} R_{\mathrm{s}}^2 - 2\omega_{\mathrm{k}}R_{\mathrm{s}}L_{\mathrm{m}} + \omega_{\mathrm{k}}^2[(L_{\mathrm{s}}^d)^2 + L_{\mathrm{m}}^2], & \omega_{\mathrm{k}}R_{\mathrm{s}}(L_{\mathrm{s}}^d - L_{\mathrm{s}}^q) + \omega_{\mathrm{k}}^2 L_{\mathrm{m}}(L_{\mathrm{s}}^d + L_{\mathrm{s}}^q) \\ \omega_{\mathrm{k}}R_{\mathrm{s}}(L_{\mathrm{s}}^d - L_{\mathrm{s}}^q) + \omega_{\mathrm{k}}^2 L_{\mathrm{m}}(L_{\mathrm{s}}^d + L_{\mathrm{s}}^q), & R_{\mathrm{s}}^2 + 2\omega_{\mathrm{k}}R_{\mathrm{s}}L_{\mathrm{m}} + \omega_{\mathrm{k}}^2[(L_{\mathrm{s}}^q)^2 + L_{\mathrm{m}}^2] \end{bmatrix} = \boldsymbol{V}(\omega_{\mathrm{k}})^\top, \\
\boldsymbol{v}(\omega_{\mathrm{k}})^\top \;&:=\; \begin{pmatrix} v_1(\omega_{\mathrm{k}}) \\ v_2(\omega_{\mathrm{k}}) \end{pmatrix} = \omega_{\mathrm{k}}\,(\boldsymbol{\psi}_{\mathrm{pm}}^k)^\top\big(\omega_{\mathrm{k}}\boldsymbol{L}_{\mathrm{s}}^k + R_{\mathrm{s}}\boldsymbol{J}^\top\big) \overset{(2)}{=} \begin{cases} \begin{pmatrix} \omega_{\mathrm{k}}^2 L_{\mathrm{s}}^d \psi_{\mathrm{pm}} \\ \omega_{\mathrm{k}}\psi_{\mathrm{pm}}(R_{\mathrm{s}} + \omega_{\mathrm{k}}L_{\mathrm{m}}) \end{pmatrix}^\top, & \text{for PMSM or PME-RSM} \\ \begin{pmatrix} -\omega_{\mathrm{k}}\psi_{\mathrm{pm}}(R_{\mathrm{s}} - \omega_{\mathrm{k}}L_{\mathrm{m}}) \\ \omega_{\mathrm{k}}^2 L_{\mathrm{s}}^q \psi_{\mathrm{pm}} \end{pmatrix}^\top, & \text{for PMA-RSM} \\ (0,\,0), & \text{for RSM} \end{cases} \\
\nu(\omega_{\mathrm{k}}, \hat{u}_{\max}) \;&:=\; \omega_{\mathrm{k}}^2\,(\boldsymbol{\psi}_{\mathrm{pm}}^k)^\top\boldsymbol{J}^\top\boldsymbol{J}\boldsymbol{\psi}_{\mathrm{pm}}^k - \hat{u}_{\max}^2 \overset{(2)}{=} \begin{cases} \omega_{\mathrm{k}}^2\,\psi_{\mathrm{pm}}^2 - \hat{u}_{\max}^2, & \text{for PMSM, PME-RSM and PMA-RSM} \\ -\hat{u}_{\max}^2, & \text{for RSM,} \end{cases}
\end{aligned}\right\} \tag{15}$$

the voltage constraint in (14) (or in (5)) can be expressed by

$$\|\boldsymbol{u}_{\mathrm{s}}^k\|^2 - \hat{u}_{\max}^2 \overset{(14),(15)}{=} (\boldsymbol{i}_{\mathrm{s}}^k)^\top\boldsymbol{V}(\omega_{\mathrm{k}})\,\boldsymbol{i}_{\mathrm{s}}^k + 2\,\boldsymbol{v}(\omega_{\mathrm{k}})^\top\boldsymbol{i}_{\mathrm{s}}^k + \nu(\omega_{\mathrm{k}}, \hat{u}_{\max}) \leq 0.$$

Note that each summand in (12) has the unit $\mathrm{VAs} = \mathrm{Nm}$.Finally, the voltage constraint (14) can be written implicitly as quadric surface defined by

$$\mathbb{V}(\omega_{\mathrm{k}}, \hat{u}_{\max}) := \left\{\, \boldsymbol{i}_{\mathrm{s}}^k \in \mathbb{R}^2 \;\middle|\; (\boldsymbol{i}_{\mathrm{s}}^k)^\top\boldsymbol{V}(\omega_{\mathrm{k}})\,\boldsymbol{i}_{\mathrm{s}}^k + 2\,\boldsymbol{v}(\omega_{\mathrm{k}})^\top\boldsymbol{i}_{\mathrm{s}}^k + \nu(\omega_{\mathrm{k}}, \hat{u}_{\max}) \leq 0 \right\}, \tag{16}$$

which describes the *voltage elliptical area*. Each summand in (16) has the unit $\mathrm{V}^2$. The *voltage ellipse* is given by

$$\boxed{\partial\mathbb{V}(\omega_{\mathrm{k}}, \hat{u}_{\max}) := \left\{\, \boldsymbol{i}_{\mathrm{s}}^k \in \mathbb{R}^2 \;\middle|\; \underbrace{(\boldsymbol{i}_{\mathrm{s}}^k)^\top\boldsymbol{V}(\omega_{\mathrm{k}})\,\boldsymbol{i}_{\mathrm{s}}^k + 2\,\boldsymbol{v}(\omega_{\mathrm{k}})^\top\boldsymbol{i}_{\mathrm{s}}^k + \nu(\omega_{\mathrm{k}}, \hat{u}_{\max})}_{=:Q_{\partial\mathbb{V}}(\boldsymbol{i}_{\mathrm{s}}^k, \omega_{\mathrm{k}}, \hat{u}_{\max})} = 0 \right\},} \tag{17}$$

i.e. the boundary of (16) (see — line in Fig. 3). The quadric $Q_{\partial\mathbb{V}}(\boldsymbol{i}_{\mathrm{s}}^k, \omega_{\mathrm{k}}, \hat{u}_{\max})$ of the voltage ellipse depends on currents $\boldsymbol{i}_{\mathrm{s}}^k = (i_{\mathrm{s}}^d, i_{\mathrm{s}}^q)^\top$, angular velocity $\omega_{\mathrm{k}}$ and voltage limit $\hat{u}_{\max}$.

**Remark III.2** (Speed dependency of the voltage ellipse and symmetry of $\boldsymbol{V}(\omega_{\mathrm{k}})$)**.** *Note that* $\boldsymbol{V}(\omega_{\mathrm{k}})$*,* $\boldsymbol{v}(\omega_{\mathrm{k}})$ *and* $\nu(\omega_{\mathrm{k}}, \hat{u}_{\max})$ *explicitly depend on the electrical angular velocity* $\omega_{\rm k}$*. Moreover, note that, the matrix* $\boldsymbol{V}(\omega_{\rm k}) = \boldsymbol{V}(\omega_{\rm k})^\top \in \mathbb{R}^{2\times 2}$ *is* indeed *symmetric for all* $\omega_{\rm k} \in \mathbb{R}$*, since all its sub-matrices are symmetric, respectively, i.e.* $(R_{\rm s}\boldsymbol{I}_2)^\top = R_{\rm s}\boldsymbol{I}_2$, $\big(\boldsymbol{J}\boldsymbol{L}_{\rm s}^k + \boldsymbol{L}_{\rm s}^k\boldsymbol{J}^\top\big)^\top = (\boldsymbol{J}\boldsymbol{L}_{\rm s}^k)^\top + (\boldsymbol{L}_{\rm s}^k\boldsymbol{J}^\top)^\top = \boldsymbol{L}_{\rm s}^k\boldsymbol{J}^\top + \boldsymbol{J}\boldsymbol{L}_{\rm s}^k = \begin{bmatrix} -2L_{\rm m} & L_{\rm s}^d - L_{\rm s}^q \\ L_{\rm s}^d - L_{\rm s}^q & 2L_{\rm m}\end{bmatrix}$ *and* $((\boldsymbol{L}_{\rm s}^k)^2)^\top = ((\boldsymbol{L}_{\rm s}^k)^\top\boldsymbol{L}_{\rm s}^k)^\top = (\boldsymbol{L}_{\rm s}^k)^\top\boldsymbol{L}_{\rm s}^k = (\boldsymbol{L}_{\rm s}^k)^2 = \begin{bmatrix} (L_{\rm s}^d)^2 + L_{\rm m}^2 & L_{\rm m}(L_{\rm s}^d + L_{\rm s}^q) \\ L_{\rm m}(L_{\rm s}^d + L_{\rm s}^q) & (L_{\rm s}^q)^2 + L_{\rm m}^2\end{bmatrix}$.

*3) Current circular area (reformulation of the current constraint in* (5)*):* The current constraint in (5) can also be expressed implicitly as quadric as follows

$$\mathbb{I}(\hat{\imath}_{\max}) := \left\{\, \boldsymbol{i}_{\rm s}^k \in \mathbb{R}^2 \;\middle|\; (\boldsymbol{i}_{\rm s}^k)^\top \boldsymbol{I}_2 \boldsymbol{i}_{\rm s}^k - \hat{\imath}_{\max}^2 \le 0 \,\right\} \tag{18}$$

which describes the admissible maximum current circular area: The magnitude of the stator current vector must not exceed the current limit $\hat{\imath}_{\max}$. The maximum current circle (see —— line in Fig. 3), i.e. the boundary of (18), is given by

$$\boxed{\partial\mathbb{I}(\hat{\imath}_{\max}) := \left\{\, \boldsymbol{i}_{\rm s}^k \in \mathbb{R}^2 \;\middle|\; (\boldsymbol{i}_{\rm s}^k)^\top \boldsymbol{I}_2 \boldsymbol{i}_{\rm s}^k - \hat{\imath}_{\max}^2 = 0 \,\right\}.} \tag{19}$$

**Remark III.3** (Explicit expression for the maximum current circle)**.** *The current circle is given by* $i_{\rm s}^q = \pm\sqrt{\hat{\imath}_{\max}^2 - (i_{\rm s}^d)^2}$.

*4) Flux norm:* To operate the machine with the MTPF strategy, the (squared) norm of the flux linkage is minimized. It can also be expressed as quadric as follows

$$\|\boldsymbol{\psi}_{\rm pm}^k\|^2 \overset{(2)}{=} (\boldsymbol{L}_{\rm s}^k\boldsymbol{i}_{\rm s}^k + \boldsymbol{\psi}_{\rm pm}^k)^\top(\boldsymbol{L}_{\rm s}^k\boldsymbol{i}_{\rm s}^k + \boldsymbol{\psi}_{\rm pm}^k) =: \; (\boldsymbol{i}_{\rm s}^k)^\top \boldsymbol{F}\boldsymbol{i}_{\rm s}^k + 2\boldsymbol{f}\boldsymbol{i}_{\rm s}^k + \phi, \tag{20}$$

where

$$\left.\begin{aligned}
\boldsymbol{F} &:= (\boldsymbol{L}_{\rm s}^k)^2 = \begin{bmatrix} (L_{\rm s}^d)^2 + L_{\rm m}^2 & L_{\rm m}(L_{\rm s}^d + L_{\rm s}^q) \\ L_{\rm m}(L_{\rm s}^d + L_{\rm s}^q) & (L_{\rm s}^q)^2 + L_{\rm m}^2 \end{bmatrix} = \boldsymbol{F}^\top, \\
\boldsymbol{f} &:= \boldsymbol{L}_{\rm s}^k\boldsymbol{\psi}_{\rm pm}^k \overset{(2)}{=} \begin{cases} \psi_{\rm pm}(L_{\rm s}^d,\, L_{\rm m})^\top, & \text{for PMSM and PME-RSM} \\ \psi_{\rm pm}(L_{\rm m},\, L_{\rm s}^q)^\top, & \text{for PMA-RSM} \\ (0,\, 0)^\top, & \text{for RSM,} \end{cases} \\
\phi &:= (\boldsymbol{\psi}_{\rm pm}^k)^\top\boldsymbol{\psi}_{\rm pm}^k = \|\boldsymbol{\psi}_{\rm pm}^k\|^2 = (\psi_{\rm pm}^d)^2 + (\psi_{\rm pm}^q)^2 \overset{(2)}{=} \begin{cases} \psi_{\rm pm}^2, & \text{for PMSM, PME-RSM and PMA-RSM} \\ 0, & \text{for RSM} \end{cases}
\end{aligned}\right\} \tag{21}$$

are the corresponding matrix, vector and scalar of the flux linkage quadric. Each summand in (20) has the unit $(\mathrm{Vs})^2$.

## IV. Operation strategies

In this section, the optimal operation strategies *Maximum Torque per Current (MTPC)*, *Maximum Current (MC)*, *Field Weakening (FW)*, *Maximum Torque per Voltage (MTPV)* and *Maximum Torque per Flux (MTPF)* are discussed in more detail and the analytical solutions of the respective reference currents are presented. Finally, the operation strategies are explained based on the visualization of the current loci (see Fig. 3). The significant impact of neglecting stator resistance and mutual inductance on the efficiency of the machine is illustrated (see Fig. 4) and discussed.

### *A. Maximum-Torque-per-Current (MTPC) hyperbola (considering* $L_{\rm m}$*)*

For low speeds, the voltage constraint in (5) is *not* critical. The current constraint in (5) and the minimization of (copper) losses dominate the operation of the machine which requires the use of the MTPC strategy (or mostly called Maximum-Torque-per-Ampere (MTPA) [34, Sec. 16.7.1] or [12], [17], [26]). The MTPC optimization problem is formulated as follows

$$\max_{\boldsymbol{i}_{\rm s}^k \in \mathbb{S}} -\|\boldsymbol{i}_{\rm s}^k\|^2 \quad \text{s.t.} \quad m_{\rm m}(\boldsymbol{i}_{\rm s}^k) = (\boldsymbol{i}_{\rm s}^k)^\top \boldsymbol{T}\boldsymbol{i}_{\rm s}^k + 2\boldsymbol{t}^\top(\boldsymbol{i}_{\rm s}^k) \overset{!}{=} m_{\rm m,ref} \big(\overset{(9)}{=} -\tau(m_{\rm m,ref})\big) \quad \text{with} \quad \mathbb{S} := \mathbb{V}(\omega_{\rm k}, \hat{u}_{\max}) \cap \mathbb{I}(\hat{\imath}_{\max}). \tag{22}$$

The admissible solution set $\mathbb{S}$ is the intersection of voltage elliptical area $\mathbb{V}(\omega_{\rm k}, \hat{u}_{\max})$ and current circular area $\mathbb{I}(\hat{\imath}_{\max})$. Its solution, the MTPC hyperbola (see —— line in Fig. 3), is given by the quadric

$$\boxed{\mathbb{MTPC} := \left\{\, \boldsymbol{i}_{\rm s}^k \in \mathbb{R}^2 \;\middle|\; (\boldsymbol{i}_{\rm s}^k)^\top \boldsymbol{M}_{\rm C}\, \boldsymbol{i}_{\rm s}^k + 2\,\boldsymbol{m}_{\rm C}^\top \boldsymbol{i}_{\rm s}^k = 0 \right\}} \tag{23}$$

where

$$
\left.
\begin{aligned}
\boldsymbol{M}_{\mathrm{C}} &:= \tfrac{3}{2}n_{\mathrm{p}} \begin{bmatrix} \frac{L_{\mathrm{s}}^d - L_{\mathrm{s}}^q}{2} & L_{\mathrm{m}} \\ L_{\mathrm{m}}, & -\frac{L_{\mathrm{s}}^d - L_{\mathrm{s}}^q}{2} \end{bmatrix} = \boldsymbol{M}_{\mathrm{C}}^\top \qquad \text{and} \\
\boldsymbol{m}_{\mathrm{C}} &:= \tfrac{3}{2}n_{\mathrm{p}} \begin{pmatrix} \frac{\psi_{\mathrm{pm}}^d}{4} \\ \frac{\psi_{\mathrm{pm}}^q}{4} \end{pmatrix} \overset{(2)}{=} \begin{cases} \tfrac{3}{2}n_{\mathrm{p}}\,(\psi_{\mathrm{pm}},\, 0)^\top, & \text{for PMSM and PME-RSM} \\ \tfrac{3}{2}n_{\mathrm{p}}\,(0,\, \psi_{\mathrm{pm}})^\top, & \text{for PMA-RSM} \\ (0,\, 0)^\top, & \text{for RSM.} \end{cases}
\end{aligned}
\right\} \tag{24}
$$

The derivation of the implicit form (23) with its parametrization (24) is presented in Appendix B. Note that the presented derivation can also be applied to obtain the implicit forms of MC, FW, MTPV and MTPF.

**Remark IV.1** (Explicit expression for the MTPC hyperbola)**.** *Depending on the parameters $L_{\mathrm{s}}^d$, $L_{\mathrm{s}}^q$ and $L_{\mathrm{m}}$, the hyperbola can be expressed explicitly by*

$$
\mathbb{MTPC}(i_{\mathrm{s}}^d) = \begin{cases} \dfrac{2L_{\mathrm{m}} i_{\mathrm{s}}^d + \frac{\psi_{\mathrm{pm}}^q}{2} \pm \sqrt{[(L_{\mathrm{s}}^d - L_{\mathrm{s}}^q)^2 + 4L_{\mathrm{m}}^2](i_{\mathrm{s}}^d)^2 + (2\psi_{\mathrm{pm}}^q L_{\mathrm{m}} + (L_{\mathrm{s}}^d - L_{\mathrm{s}}^q)\psi_{\mathrm{pm}}^d) i_{\mathrm{s}}^d + \frac{(\psi_{\mathrm{pm}}^q)^2}{4}}}{L_{\mathrm{s}}^d - L_{\mathrm{s}}^q} & \textit{for } L_{\mathrm{s}}^d \neq L_{\mathrm{s}}^q \\ -(i_{\mathrm{s}}^d = 0), & \textit{for } L_{\mathrm{s}}^d = L_{\mathrm{s}}^q, \end{cases} \tag{25}
$$

*which holds only for $i_{\mathrm{s}}^d \leq 0$ and $L_{\mathrm{s}}^q \geq L_{\mathrm{s}}^d$. For $i_{\mathrm{s}}^d > 0$ and/or $L_{\mathrm{s}}^d = L_{\mathrm{s}}^q$, another explicit expression has to be found. The implicit form* (23) *with* (24) *holds in general (a significant advantage obviating the need of case studies). The mathematical derivation of the explicit expression of a general quadric is explained in Appendix C.*

**Remark IV.2** (MTPC versus MTPA)**.** *In most publications, the MTPC strategy is called Maximum Torque Per Ampere (MTPA). From a physical point of view, the use of physical quantities in the terminology (like torque and current) seems more appropriate than a mixture of quantity and unit (like torque and Ampere). Therefore, in this paper, the terminology MTPC will be adopted instead of MTPA (following the publications [30], [39], [40]).*

### *B. Maximum current (MC)*

To operate the machine at its current limit for increasing angular velocities, the MC strategy is used where the maximally feasible torque should be produced, i.e.

$$
\boxed{\max_{\boldsymbol{i}_{\mathrm{s}}^k \in \mathbb{S}} \operatorname{sign}(m_{\mathrm{m,ref}}) m_{\mathrm{m}}(\boldsymbol{i}_{\mathrm{s}}^k) \;\; \text{with} \;\; \mathbb{S} := \partial\mathbb{I}(\hat{\imath}_{\max}) \cap \mathbb{V}(\omega_{\mathrm{k}}, \hat{u}_{\max}).}
$$

The optimal reference currents are obtained as the intersection points of the current circle $\partial\mathbb{I}(\hat{\imath}_{\max})$ and the voltage ellipse $\partial\mathbb{V}(\omega_{\mathrm{k}}, \hat{u}_{\max})$. Hence, the reference current vectors are element of the following set

$$
\boxed{\begin{aligned} \mathbb{MC}(\omega_{\mathrm{k}}, \hat{u}_{\max}, \hat{\imath}_{\max}) &:= \partial\mathbb{I}(\hat{\imath}_{\max}) \cap \partial\mathbb{V}(\omega_{\mathrm{k}}, \hat{u}_{\max}) \\ &= \Big\{ \boldsymbol{i}_{\mathrm{s}}^k \in \mathbb{R}^2 \;\Big|\; (\boldsymbol{i}_{\mathrm{s}}^k)^\top \boldsymbol{I}_2\, \boldsymbol{i}_{\mathrm{s}}^k - \hat{\imath}_{\max}^2 = 0 \;\; \wedge \\ &\qquad (\boldsymbol{i}_{\mathrm{s}}^k)^\top \boldsymbol{V}(\omega_{\mathrm{k}}) \boldsymbol{i}_{\mathrm{s}}^k + 2\, \boldsymbol{v}(\omega_{\mathrm{k}})^\top \boldsymbol{i}_{\mathrm{s}}^k + \nu(\omega_{\mathrm{k}}, \hat{u}_{\max}) \Big\} \end{aligned}} \tag{26}
$$

with $\boldsymbol{V}$, $\boldsymbol{v}$ and $\nu$ as in (15). An algorithm to compute these intersections points *analytically* is presented in Appendix D.

### *C. Field weakening (FW)*

For a feasible torque below rated machine torque and angular velocities higher than a certain feasible MTPC velocity, the machine is operated in FW. The optimization problem for FW is identical to the optimization problem for MTPC as in (22). Due to a smaller feasible set $\mathbb{S} := \mathbb{V}(\omega_{\mathrm{k}}, \hat{u}_{\max}) \cap \mathbb{I}(\hat{\imath}_{\max})$, the optimal reference currents are obtained by the intersection of the (feasible) torque hyperbola $\mathbb{T}(m_{\mathrm{m,ref}})$ and the voltage ellipse $\partial\mathbb{V}(\omega_{\mathrm{k}}, \hat{u}_{\max})$ and, hence, the reference current vector is element of the following set

$$
\boxed{\begin{aligned} \mathbb{FW}(m_{\mathrm{m,ref}}, \omega_{\mathrm{k}}, \hat{u}_{\max}) &:= \mathbb{T}(m_{\mathrm{m,ref}}) \cap \partial\mathbb{V}(\omega_{\mathrm{k}}, \hat{u}_{\max}) \\ &= \Big\{ \boldsymbol{i}_{\mathrm{s}}^k \in \mathbb{R}^2 \;\Big|\; (\boldsymbol{i}_{\mathrm{s}}^k)^\top \boldsymbol{T}\, \boldsymbol{i}_{\mathrm{s}}^k + 2\, \boldsymbol{t}^\top \boldsymbol{i}_{\mathrm{s}}^k + \tau(m_{\mathrm{m,ref}}) = 0 \;\; \wedge \\ &\qquad (\boldsymbol{i}_{\mathrm{s}}^k)^\top \boldsymbol{V}(\omega_{\mathrm{k}}) \boldsymbol{i}_{\mathrm{s}}^k + 2\, \boldsymbol{v}(\omega_{\mathrm{k}})^\top \boldsymbol{i}_{\mathrm{s}}^k + \nu(\omega_{\mathrm{k}}, \hat{u}_{\max}) = 0 \Big\} \end{aligned}} \tag{27}
$$

with $\boldsymbol{T}$, $\boldsymbol{t}$ and $\tau(m_{\mathrm{m,ref}})$ as in (9) and $\boldsymbol{V}(\omega_{\mathrm{k}})$, $\boldsymbol{v}(\omega_{\mathrm{k}})$ and $\nu(\omega_{\mathrm{k}}, \hat{u}_{\max})$ as in (15). Again, the computation of the intersection points is based on the analytical algorithm presented in Appendix D.

*D. Maximum-Torque-per-Voltage (MTPV) hyperbola (considering $R_\mathrm{s}$ and $L_\mathrm{m}$)*

For speeds higher than the MTPV cut-in speed $\omega_{\mathrm{k,cut\text{-}in}}^{\mathrm{MTPV}}$ and for torques higher than or equal to the speed-dependent MTPV cut-in torque $m_{\mathrm{m,cut\text{-}in}}^{\mathrm{MTPV}}$, the voltage constraint in (5) is critical and dominates the operation of the machine. The operation strategy now is MTPV. The corresponding MPTV optimization problem is formulated as follows

$$\max_{\boldsymbol{i}_\mathrm{s}^k\in\mathbb{S}} -\|\boldsymbol{u}_\mathrm{s}^k(\boldsymbol{i}_\mathrm{s}^k)\|^2 \quad \text{s.t.} \quad m_\mathrm{m}(\boldsymbol{i}_\mathrm{s}^k)=(\boldsymbol{i}_\mathrm{s}^k)^\top\boldsymbol{T}\boldsymbol{i}_\mathrm{s}^k+2\boldsymbol{t}^\top(\boldsymbol{i}_\mathrm{s}^k)\overset{!}{=}m_\mathrm{m,ref} \quad \text{with} \quad \mathbb{S}=\mathbb{V}(\omega_\mathrm{k},\hat{u}_\mathrm{max})\cap\mathbb{I}(\hat{\imath}_\mathrm{max}). \tag{28}$$

Its solution, the MTPV hyperbola (see —— line in Fig. 3), is parametrized by the electrical angular velocity $\omega_\mathrm{k}$ and is implicitly given by the quadric

$$\boxed{\mathbb{MTPV}(\omega_\mathrm{k}):=\left\{\,\boldsymbol{i}_\mathrm{s}^k\in\mathbb{R}^2 \;\middle|\; (\boldsymbol{i}_\mathrm{s}^k)^\top\boldsymbol{M}_\mathrm{V}(\omega_\mathrm{k})\,\boldsymbol{i}_\mathrm{s}^k+2\,\boldsymbol{m}_\mathrm{V}(\omega_\mathrm{k})^\top\boldsymbol{i}_\mathrm{s}^k+\mu_\mathrm{V}(\omega_\mathrm{k})=0\right\}} \tag{29}$$

where

$$\left.\begin{aligned}
\boldsymbol{M}_\mathrm{V}(\omega_\mathrm{k}) &:= \begin{bmatrix} m_\mathrm{V}^{11}(\omega_\mathrm{k}), & m_\mathrm{V}^{12}(\omega_\mathrm{k})\\ m_\mathrm{V}^{12}(\omega_\mathrm{k}), & m_\mathrm{V}^{22}(\omega_\mathrm{k})\end{bmatrix}=\boldsymbol{M}_\mathrm{V}(\omega_\mathrm{k})^\top= \\
&= \tfrac{3}{2}n_\mathrm{p}\begin{bmatrix} \begin{array}{c}\omega_\mathrm{k}^2L_\mathrm{m}^2(L_\mathrm{s}^d+L_\mathrm{s}^q)\\ +\frac{L_\mathrm{s}^d-L_\mathrm{s}^q}{2}\Big(R_\mathrm{s}^2+\omega_\mathrm{k}^2\big[(L_\mathrm{s}^d)^2+L_\mathrm{m}^2\big]\Big)\end{array}, & \frac{L_\mathrm{m}}{2}\Big(2R_\mathrm{s}^2+\omega_\mathrm{k}^2\big[(L_\mathrm{s}^d)^2+(L_\mathrm{s}^q)^2+2L_\mathrm{m}^2\big]\Big) \\[2ex] \frac{L_\mathrm{m}}{2}\Big(2R_\mathrm{s}^2+\omega_\mathrm{k}^2\big[(L_\mathrm{s}^d)^2+(L_\mathrm{s}^q)^2+2L_\mathrm{m}^2\big]\Big), & \begin{array}{c}\omega_\mathrm{k}^2L_\mathrm{m}^2(L_\mathrm{s}^d+L_\mathrm{s}^q)\\ -\frac{L_\mathrm{s}^d-L_\mathrm{s}^q}{2}\Big(R_\mathrm{s}^2+\omega_\mathrm{k}^2\big[(L_\mathrm{s}^q)^2+L_\mathrm{m}^2\big]\Big)\end{array}\end{bmatrix}, \\
\boldsymbol{m}_\mathrm{V}(\omega_\mathrm{k}) &:= \begin{pmatrix} m_\mathrm{V}^1(\omega_\mathrm{k})\\ m_\mathrm{V}^2(\omega_\mathrm{k})\end{pmatrix}=\tfrac{3}{2}n_\mathrm{p}\begin{pmatrix}\Big(R_\mathrm{s}^2+\omega_\mathrm{k}^2\big[2(L_\mathrm{s}^d)^2-L_\mathrm{s}^qL_\mathrm{s}^d+3L_\mathrm{m}^2\big]\Big)\frac{\psi_\mathrm{pm}^d}{4}+\omega_\mathrm{k}^2L_\mathrm{m}(L_\mathrm{s}^d+L_\mathrm{s}^q)\frac{\psi_\mathrm{pm}^q}{2}\\ \omega_\mathrm{k}^2L_\mathrm{m}(L_\mathrm{s}^d+L_\mathrm{s}^q)\frac{\psi_\mathrm{pm}^d}{2}+\Big(R_\mathrm{s}^2+\omega_\mathrm{k}^2\big[2(L_\mathrm{s}^q)^2-L_\mathrm{s}^qL_\mathrm{s}^d+3L_\mathrm{m}^2\big]\Big)\frac{\psi_\mathrm{pm}^q}{4}\end{pmatrix} \quad \text{and} \\
\mu_\mathrm{V}(\omega_\mathrm{k}) &:= \tfrac{3}{4}n_\mathrm{p}\omega_\mathrm{k}^2\Big[L_\mathrm{s}^d(\psi_\mathrm{pm}^d)^2+2L_\mathrm{m}\psi_\mathrm{pm}^d\psi_\mathrm{pm}^q+L_\mathrm{s}^q(\psi_\mathrm{pm}^q)^2\Big]\overset{(2)}{=}\begin{cases}\frac{3}{4}n_\mathrm{p}\omega_\mathrm{k}^2L_\mathrm{s}^d\psi_\mathrm{pm}^2, & \text{for PMSM and PME-RSM}\\ \frac{3}{4}n_\mathrm{p}\omega_\mathrm{k}^2L_\mathrm{s}^q\psi_\mathrm{pm}^2, & \text{for PMA-RSM}\\ 0, & \text{for RSM.}\end{cases}
\end{aligned}\right\} \tag{30}$$

Obviously, since $\boldsymbol{M}_\mathrm{V}(\omega_\mathrm{k})$, $\boldsymbol{m}_\mathrm{V}(\omega_\mathrm{k})$ and $\mu_\mathrm{V}(\omega_\mathrm{k})$ *all* depend on the angular velocity $\omega_\mathrm{k}$, the MTPV hyperbola is moving in the $(i_\mathrm{s}^d, i_\mathrm{s}^q)$-plane (see —— lines in Fig. 3(a), (b) & (c)). Again, the derivation of the implicit form (29) with its parametrization (30) follows Appendix B.

**Remark IV.3** (Explicit expression for the MTPV hyperbola)**.** *The explicit solution of the MTPV hyperbola is given by (see Appendix C)*

$$\mathbb{MTPV}(i_\mathrm{s}^d,\omega_\mathrm{k})=-\frac{m_\mathrm{V}^{12}\,i_\mathrm{s}^d+m_\mathrm{V}^2}{m_\mathrm{V}^{22}}\pm\frac{\sqrt{\Big(m_\mathrm{V}^{12}\,i_\mathrm{s}^d+m_\mathrm{V}^2\Big)^2-m_\mathrm{V}^{22}\Big(m_\mathrm{V}^{11}\,(i_\mathrm{s}^d)^2+2m_\mathrm{V}^1\,i_\mathrm{s}^d+\mu_\mathrm{V}\Big)}}{m_\mathrm{V}^{22}}, \tag{31}$$

*where $m_\mathrm{V}^{11}=m_\mathrm{V}^{11}(\omega_\mathrm{k})$, $m_\mathrm{V}^{12}=m_\mathrm{V}^{12}(\omega_\mathrm{k})$, $m_\mathrm{V}^{22}=m_\mathrm{V}^{22}(\omega_\mathrm{k})$, $m_\mathrm{V}^1=m_\mathrm{V}^1(\omega_\mathrm{k})$, $m_\mathrm{V}^2=m_\mathrm{V}^2(\omega_\mathrm{k})$ and $\mu_\mathrm{V}=\mu_\mathrm{V}(\omega_\mathrm{k})$ are as in* (30)*. Note that $m_\mathrm{V}^{22}(0)=-\frac{L_\mathrm{s}^d-L_\mathrm{s}^q}{2}R_\mathrm{s}^2\neq 0$ and, hence, is non-zero for all $\omega_\mathrm{k}$ and all $L_\mathrm{s}^d\neq L_\mathrm{s}^q$.*

**Remark IV.4** (MTPV hyperbola without stator resistance)**.** *Note that the MTPV hyperbola without stator resistance can be obtained from* (30) *by setting $R_\mathrm{s}=0$. This was already shown in [30].*

*E. Maximum-Torque-per-Flux (MTPF) hyperbola (considering $L_\mathrm{m}$)*

For high speeds, an alternative to the MTPV strategy is the MTPF strategy. Nevertheless, it yields a reference current vector with larger magnitude than that obtained from MTPV; hence, the MTPV strategy should be used instead (see also Remark IV.6). The MTPF optimization problem can be formulated as follows

$$\max_{\boldsymbol{i}_\mathrm{s}^k\in\mathbb{S}} -\|\boldsymbol{\psi}_\mathrm{s}^k(\boldsymbol{i}_\mathrm{s}^k)\|^2 \quad \text{s.t.} \quad m_\mathrm{m}(\boldsymbol{i}_\mathrm{s}^k)=(\boldsymbol{i}_\mathrm{s}^k)^\top\boldsymbol{T}\boldsymbol{i}_\mathrm{s}^k+2\boldsymbol{t}^\top(\boldsymbol{i}_\mathrm{s}^k)\overset{!}{=}m_\mathrm{m,ref} \quad \text{with} \quad \mathbb{S}:=\mathbb{V}(\omega_\mathrm{k},\hat{u}_\mathrm{max})\cap\mathbb{I}(\hat{\imath}_\mathrm{max}). \tag{32}$$

Its solution, the MTPF hyperbola (see —— line in Fig. 3) is implicitly given by the quadric

$$\boxed{\mathbb{MTPF}:=\left\{\,\boldsymbol{i}_\mathrm{s}^k\in\mathbb{R}^2 \;\middle|\; (\boldsymbol{i}_\mathrm{s}^k)^\top\boldsymbol{M}_\mathrm{F}\,\boldsymbol{i}_\mathrm{s}^k+2\,\boldsymbol{m}_\mathrm{F}^\top\boldsymbol{i}_\mathrm{s}^k+\mu_\mathrm{F}=0\right\},} \tag{33}$$

which does *not* depend on the angular velocity $\omega_k$ (in contrast to the MTPV hyperbola), since

$$\left.\begin{aligned}
\boldsymbol{M}_F &:= \begin{bmatrix} m_F^{11}, & m_F^{12} \\ m_F^{12}, & m_F^{22} \end{bmatrix} = \boldsymbol{M}_F^\top = \\
&= \tfrac{3}{2}n_p \begin{bmatrix} \frac{L_s^d - L_s^q}{2}\big((L_s^d)^2 + L_m^2\big) + L_m^2\big(L_s^d + L_s^q\big), & \frac{L_m}{2}\big((L_s^d)^2 + (L_s^q)^2 + 2L_m^2\big) \\ \frac{L_m}{2}\big((L_s^d)^2 + (L_s^q)^2 + 2L_m^2\big), & -\frac{L_s^d - L_s^q}{2}\big((L_s^q)^2 + L_m^2\big) + L_m^2\big(L_s^d + L_s^q\big) \end{bmatrix}, \\
\boldsymbol{m}_F &:= \begin{pmatrix} m_F^1 \\ m_F^2 \end{pmatrix} = \tfrac{3}{2}n_p \begin{pmatrix} \Big(2(L_s^d)^2 - L_s^d L_s^q + 3L_m^2\Big)\frac{\psi_{pm}^d}{4} + L_m(L_s^d + L_s^q)\frac{\psi_{pm}^q}{2} \\ L_m(L_s^d + L_s^q)\frac{\psi_{pm}^d}{2} + \Big(2(L_s^q)^2 - L_s^d L_s^q + 3L_m^2\Big)\frac{\psi_{pm}^q}{4} \end{pmatrix} \quad \text{and} \\
\mu_F &:= \tfrac{3}{4}n_p\Big[L_s^d(\psi_{pm}^d)^2 + 2L_m\psi_{pm}^d\psi_{pm}^q + L_s^q(\psi_{pm}^q)^2\Big] \overset{(2)}{=} \begin{cases} \frac{3}{4}n_p L_s^d(\psi_{pm})^2, & \text{for PMSM and PME-RSM} \\ \frac{3}{4}n_p L_s^q(\psi_{pm})^2, & \text{for PMA-RSM} \\ 0, & \text{for RSM,} \end{cases}
\end{aligned}\right\} \tag{34}$$

do *not* depend on the electrical angular velocity $\omega_k$, respectively. As before for MTPC and MTPV, the derivation of the implicit form (33) with its parametrization (34) is based on Appendix B.

**Remark IV.5** (Explicit expression for the MTPF hyperbola)**.** *The explicit solution of the MTPF hyperbola is given by (see Appendix C)*

$$\mathbb{MTPF}(i_s^d) = -\frac{m_F^{12}\, i_s^d + m_F^2}{m_F^{22}} \pm \frac{\sqrt{\Big(m_F^{12}\, i_s^d + m_F^2\Big)^2 - m_F^{22}\Big(m_F^{11}\,(i_s^d)^2 + 2m_F^1\, i_s^d + \mu_F\Big)}}{m_F^{22}}, \tag{35}$$

*where* $m_F^{11}$, $m_F^{12}$, $m_F^{22}$, $m_F^1$, $m_F^2$ *and* $\mu_F$ *are as in* (34)*. Note that* $m_F^{22} = \frac{3}{2}n_p\big[-\frac{L_s^d - L_s^q}{2}\big((L_s^q)^2 + L_m^2\big) + L_m^2\big(L_s^d + L_s^q\big)\big] \neq 0.$

**Remark IV.6** (Convergence of the MTPV hyperbola to the MTPF hyperbola)**.** *For very large electrical angular velocities* $|\omega_k| \gg 1$ *or very small stator resistances* $R_s \ll 1\,\Omega$*, the MTPV hyperbola converges to the shape of the MTPF hyperbola, since, either for* $\omega_k \to \infty$ *or for* $R_s = 0$*, the following holds*

$$(\boldsymbol{i}_s^k)^\top \boldsymbol{M}_F\, \boldsymbol{i}_s^k + 2\,\boldsymbol{m}_F^\top \boldsymbol{i}_s^k + \mu_F = 0 = \begin{cases} \lim\limits_{\omega_k \to \infty} (\boldsymbol{i}_s^k)^\top \frac{\boldsymbol{M}_V(\omega_k)}{\omega_k^2}\, \boldsymbol{i}_s^k + 2\,\frac{\boldsymbol{m}_V(\omega_k)^\top}{\omega_k^2}\boldsymbol{i}_s^k + \frac{\mu_V(\omega_k)}{\omega_k^2}, & R_s \neq 0 \\ (\boldsymbol{i}_s^k)^\top \frac{\boldsymbol{M}_V(\omega_k)}{\omega_k^2}\, \boldsymbol{i}_s^k + 2\,\frac{\boldsymbol{m}_V(\omega_k)^\top}{\omega_k^2}\boldsymbol{i}_s^k + \frac{\mu_V(\omega_k)}{\omega_k^2}, & R_s = 0. \end{cases}$$

*Concluding,* only *for very large speeds or very small values of the stator resistance, both strategies are similar. In general, MTPF and MTPV hyperbola are* different *solutions to different optimization problems and give different optimal reference currents (see Fig. 3c).*

### *F. Analytical solutions of the optimal reference current vector for MTPC, MC, FW, MTPV and MTPF*

As soon as, the implicit expressions (quadrics)

(i) for the constraints (i.e. voltage ellipse $\mathbb{V}(\hat{u}_{max}, \omega_k)$, current circle $\mathbb{I}(\hat{\imath}_{max})$, and torque hyperbola $\mathbb{T}(m_{m,ref})$), and
(ii) for the operation strategies MTPC, MTPV, and MTPF (i.e. $\mathbb{MTPC}$, $\mathbb{MTPV}(\omega_k)$ and $\mathbb{MTPF}$ hyperbola, respectively) are derived, the optimal reference currents $\boldsymbol{i}_{s,ref}^k = \boldsymbol{i}_{s,ref}^{k,X}$ for each operation strategy X $\in$ {MTPC, MC, FW, MTPV, MTPF} are obtained by intersecting the respective quadrics (following the general approach presented in Appendix D).

In Tab. I, for each operation strategy, the analytical expression for the optimal current reference vector and the used computation method (algorithm) are listed in compact form. In all cases, $\lambda^\star$ and $\gamma^\star$ are the optimal Lagrangian multipliers which represent one of the (real) roots of the polynomial (45) and (67), respectively. The four roots can be computed analytically by the algorithm presented in Appendix A3 (Euler's solution [41]).

**Remark IV.7** (Alternative computation of optimal reference currents for MTPC)**.** *Note that, alternatively, by using the algorithm discussed in Appendix D, the optimal current reference vectors for MTPC can also be obtained by computing the intersection points of torque hyperbola* (12) *and MTPC hyperbola* (23) *(see also Tab. I).*

**Remark IV.8** (Optimal reference currents for reluctance synchronous machines (RSMs))**.** *The analytical solutions for RSMs can be computed in a similar way as shown in Tab. I; however, for RSMs, all quadrics simplify due to the missing permanent magnet, i.e.* $\psi_{pm}^d = \psi_{pm}^q = 0$*. The vectors* $\boldsymbol{t} = \boldsymbol{v}(\omega_k) = \boldsymbol{f} = \boldsymbol{m}_C = \boldsymbol{m}_V(\omega_k) = \boldsymbol{m}_F = \boldsymbol{0}_2$ *and scalars* $\phi = \mu_V(\omega_k) = \mu_F = 0$ *of torque hyperbola* (12)*, voltage ellipse* (17)*, flux norm* (20)*, MTPC hyperbola* (23)*, MTPV hyperbola* (29) *and MTPF hyperbola* (33) *become zero (see also* (9), (15), (21), (24), (30) *and* (34)*), respectively. Therefore, instead of applying case (i) of Appendix D, case (iii) of Appendix D must be considered for the computation of the intersection points of the respective quadrics (see also Tab. I).*

| strategy | current reference vector | algorithm used |
|---|---|---|
| **MTPC** | $\boldsymbol{i}_{\mathrm{s,ref}}^{k,\mathrm{MTPC}}(\lambda^\star) \overset{(50)}{:=} -\big[\lambda^\star \boldsymbol{T} - \boldsymbol{I}_2\big]^{-1}\lambda^\star \boldsymbol{t}$ | see Appendix A<br>where $\boldsymbol{A}=\boldsymbol{I}_2$, $\boldsymbol{a}=\boldsymbol{0}_2$, $\boldsymbol{B}=\boldsymbol{T}$ & $\boldsymbol{b}=\boldsymbol{t}$ |
| or | $\boldsymbol{i}_{\mathrm{s,ref}}^{k,\mathrm{MTPC}}(\gamma^\star) \overset{(69)}{:=} -2\big[\boldsymbol{D}-\gamma^\star\boldsymbol{J}\big]^{-1}\boldsymbol{d}$ | see Appendix D (case (ii) for $\mathbb{MTPC}\cap\mathbb{T}(m_{\mathrm{m,ref}})$)<br>where $\boldsymbol{A}=\boldsymbol{M}_{\mathrm{C}}$, $\boldsymbol{a}=\boldsymbol{m}_{\mathrm{C}}$, $\alpha=0$, $\boldsymbol{B}=\boldsymbol{T}$, $\boldsymbol{b}=\boldsymbol{t}$, $\beta=-m_{\mathrm{m,ref}}$,<br>$\boldsymbol{D}=\boldsymbol{A}$ and $\boldsymbol{d}=\boldsymbol{a}$. |
| for RSM | $\boldsymbol{i}_{\mathrm{s,ref}}^{k,\mathrm{MTPC}}(\gamma^\star) \overset{(69)}{:=} -2\big[\boldsymbol{D}-\gamma^\star\boldsymbol{J}\big]^{-1}\boldsymbol{d}+\boldsymbol{x}_{\mathrm{s}}$ | see Appendix D (case (iii) for $\mathbb{MTPC}\cap\mathbb{T}(m_{\mathrm{m,ref}})$)<br>where $\boldsymbol{A}=\boldsymbol{M}_{\mathrm{C}}$, $\boldsymbol{a}=\boldsymbol{m}_{\mathrm{C}}=\boldsymbol{0}_2$, $\alpha=0$, $\boldsymbol{x}_{\mathrm{s}}\neq\boldsymbol{0}_2$ (arbitrary),<br>$\boldsymbol{B}=\boldsymbol{T}$, $\boldsymbol{b}=\boldsymbol{t}=\boldsymbol{0}_2$, $\beta=-m_{\mathrm{m,ref}}$, $\boldsymbol{D}:=\Big(\frac{\boldsymbol{A}}{\alpha_{\mathrm{s}}}-\frac{\boldsymbol{B}}{\beta_{\mathrm{s}}}\Big)$,<br>$\boldsymbol{d}:=\Big(\frac{\boldsymbol{a}_{\mathrm{s}}}{\alpha_{\mathrm{s}}}-\frac{\boldsymbol{b}_{\mathrm{s}}}{\beta_{\mathrm{s}}}\Big)$ with $\boldsymbol{a}_{\mathrm{s}}$, $\alpha_{\mathrm{s}}$, $\boldsymbol{b}_{\mathrm{s}}$ and $\beta_{\mathrm{s}}$ as in (65). |
| **MC** | $\boldsymbol{i}_{\mathrm{s,ref}}^{k,\mathrm{MC}}(\gamma^\star) \overset{(69)}{:=} -2\big[\boldsymbol{D}-\gamma^\star\boldsymbol{J}\big]^{-1}\boldsymbol{d}$ | see Appendix D (case (i) for $\partial\mathbb{I}(\hat{\imath}_{\max})\cap\partial\mathbb{V}(\omega_{\mathrm{k}},\hat{u}_{\max})$)<br>where $\boldsymbol{A}=\boldsymbol{I}_2$, $\boldsymbol{a}=\boldsymbol{0}_2$, $\alpha=-\hat{\imath}_{\max}^2$,<br>$\boldsymbol{B}=\boldsymbol{V}(\omega_{\mathrm{k}})$, $\boldsymbol{b}=\boldsymbol{v}(\omega_{\mathrm{k}})$, $\beta=\nu(\omega_{\mathrm{k}},\hat{u}_{\max})$,<br>$\boldsymbol{D}=\big(\frac{\boldsymbol{A}}{\alpha}-\frac{\boldsymbol{B}}{\beta}\big)$ and $\boldsymbol{d}=\big(\frac{\boldsymbol{a}}{\alpha}-\frac{\boldsymbol{b}}{\beta}\big)$. |
| for RSM | $\boldsymbol{i}_{\mathrm{s,ref}}^{k,\mathrm{MC}}(\gamma^\star) \overset{(69)}{:=} -2\big[\boldsymbol{D}-\gamma^\star\boldsymbol{J}\big]^{-1}\boldsymbol{d}+\boldsymbol{x}_{\mathrm{s}}$ | see Appendix D (case (iii) for $\partial\mathbb{I}(\hat{\imath}_{\max})\cap\partial\mathbb{V}(\omega_{\mathrm{k}},\hat{u}_{\max})$)<br>where $\boldsymbol{A}=\boldsymbol{I}_2$, $\boldsymbol{a}=\boldsymbol{0}_2$, $\alpha=-\hat{\imath}_{\max}^2$, $\boldsymbol{x}_{\mathrm{s}}\neq\boldsymbol{0}_2$ (arbitrary),<br>$\boldsymbol{B}=\boldsymbol{V}(\omega_{\mathrm{k}})$, $\boldsymbol{b}=\boldsymbol{v}(\omega_{\mathrm{k}})=\boldsymbol{0}_2$, $\beta=\nu(\omega_{\mathrm{k}},\hat{u}_{\max})$, $\boldsymbol{D}:=\Big(\frac{\boldsymbol{A}}{\alpha_{\mathrm{s}}}-\frac{\boldsymbol{B}}{\beta_{\mathrm{s}}}\Big)$,<br>$\boldsymbol{d}:=\Big(\frac{\boldsymbol{a}_{\mathrm{s}}}{\alpha_{\mathrm{s}}}-\frac{\boldsymbol{b}_{\mathrm{s}}}{\beta_{\mathrm{s}}}\Big)$ with $\boldsymbol{a}_{\mathrm{s}}$, $\alpha_{\mathrm{s}}$, $\boldsymbol{b}_{\mathrm{s}}$ and $\beta_{\mathrm{s}}$ as in (65). |
| **FW** | $\boldsymbol{i}_{\mathrm{s,ref}}^{k,\mathrm{FW}}(\gamma^\star) \overset{(69)}{:=} -2\big[\boldsymbol{D}-\gamma^\star\boldsymbol{J}\big]^{-1}\boldsymbol{d}$ | see Appendix D (case (i) for $\partial\mathbb{T}(m_{\mathrm{m,ref}})\cap\partial\mathbb{V}(\omega_{\mathrm{k}},\hat{u}_{\max})$)<br>where $\boldsymbol{A}=\boldsymbol{T}$, $\boldsymbol{a}=\boldsymbol{t}$, $\alpha=\tau(m_{\mathrm{m,ref}})$,<br>$\boldsymbol{B}=\boldsymbol{V}(\omega_{\mathrm{k}})$, $\boldsymbol{b}=\boldsymbol{v}(\omega_{\mathrm{k}})$, $\beta=\nu(\omega_{\mathrm{k}},\hat{u}_{\max})$,<br>$\boldsymbol{D}=\big(\frac{\boldsymbol{A}}{\alpha}-\frac{\boldsymbol{B}}{\beta}\big)$ and $\boldsymbol{d}=\big(\frac{\boldsymbol{a}}{\alpha}-\frac{\boldsymbol{b}}{\beta}\big)$ |
| for RSM | $\boldsymbol{i}_{\mathrm{s,ref}}^{k,\mathrm{FW}}(\gamma^\star) \overset{(69)}{:=} -2\big[\boldsymbol{D}-\gamma^\star\boldsymbol{J}\big]^{-1}\boldsymbol{d}+\boldsymbol{x}_{\mathrm{s}}$ | see Appendix D (case (iii) for $\partial\mathbb{T}(m_{\mathrm{m,ref}})\cap\partial\mathbb{V}(\omega_{\mathrm{k}},\hat{u}_{\max})$)<br>where $\boldsymbol{A}=\boldsymbol{T}$, $\boldsymbol{a}=\boldsymbol{t}=\boldsymbol{0}_2$, $\alpha=\tau(m_{\mathrm{m,ref}})$, $\boldsymbol{x}_{\mathrm{s}}\neq\boldsymbol{0}_2$ (arbitrary),<br>$\boldsymbol{B}=\boldsymbol{V}(\omega_{\mathrm{k}})$, $\boldsymbol{b}=\boldsymbol{v}(\omega_{\mathrm{k}})=\boldsymbol{0}_2$, $\beta=\nu(\omega_{\mathrm{k}},\hat{u}_{\max})$, $\boldsymbol{D}:=\Big(\frac{\boldsymbol{A}}{\alpha_{\mathrm{s}}}-\frac{\boldsymbol{B}}{\beta_{\mathrm{s}}}\Big)$,<br>$\boldsymbol{d}:=\Big(\frac{\boldsymbol{a}_{\mathrm{s}}}{\alpha_{\mathrm{s}}}-\frac{\boldsymbol{b}_{\mathrm{s}}}{\beta_{\mathrm{s}}}\Big)$ with $\boldsymbol{a}_{\mathrm{s}}$, $\alpha_{\mathrm{s}}$, $\boldsymbol{b}_{\mathrm{s}}$ and $\beta_{\mathrm{s}}$ as in (65). |
| **MTPF** | $\boldsymbol{i}_{\mathrm{s,ref}}^{k,\mathrm{MTPF}}(\gamma^\star) \overset{(69)}{:=} -2\big[\boldsymbol{D}-\gamma^\star\boldsymbol{J}\big]^{-1}\boldsymbol{d}$ | see Appendix D (case (i) for $\mathbb{MTPF}\cap\partial\mathbb{V}(\omega_{\mathrm{k}},\hat{u}_{\max})$)<br>where $\boldsymbol{A}=\boldsymbol{M}_{\mathrm{F}}$, $\boldsymbol{a}=\boldsymbol{m}_{\mathrm{F}}$, $\alpha=\mu_{\mathrm{F}}$,<br>$\boldsymbol{B}=\boldsymbol{V}(\omega_{\mathrm{k}})$, $\boldsymbol{b}=\boldsymbol{v}(\omega_{\mathrm{k}})$, $\beta=\nu(\omega_{\mathrm{k}},\hat{u}_{\max})$,<br>$\boldsymbol{D}=\big(\frac{\boldsymbol{A}}{\alpha}-\frac{\boldsymbol{B}}{\beta}\big)$ and $\boldsymbol{d}=\big(\frac{\boldsymbol{a}}{\alpha}-\frac{\boldsymbol{b}}{\beta}\big)$. |
| for RSM | $\boldsymbol{i}_{\mathrm{s,ref}}^{k,\mathrm{MTPF}}(\gamma^\star) \overset{(69)}{:=} -2\big[\boldsymbol{D}-\gamma^\star\boldsymbol{J}\big]^{-1}\boldsymbol{d}+\boldsymbol{x}_{\mathrm{s}}$ | see Appendix D (case (iii) for $\mathbb{MTPF}\cap\partial\mathbb{V}(\omega_{\mathrm{k}},\hat{u}_{\max})$)<br>where $\boldsymbol{A}=\boldsymbol{M}_{\mathrm{F}}$, $\boldsymbol{a}=\boldsymbol{m}_{\mathrm{F}}=\boldsymbol{0}_2$, $\alpha=\mu_{\mathrm{F}}=0$, $\boldsymbol{x}_{\mathrm{s}}\neq\boldsymbol{0}_2$ (arbitrary),<br>$\boldsymbol{B}=\boldsymbol{V}(\omega_{\mathrm{k}})$, $\boldsymbol{b}=\boldsymbol{v}(\omega_{\mathrm{k}})=\boldsymbol{0}_2$, $\beta=\nu(\omega_{\mathrm{k}},\hat{u}_{\max})$, $\boldsymbol{D}:=\Big(\frac{\boldsymbol{A}}{\alpha_{\mathrm{s}}}-\frac{\boldsymbol{B}}{\beta_{\mathrm{s}}}\Big)$,<br>$\boldsymbol{d}:=\Big(\frac{\boldsymbol{a}_{\mathrm{s}}}{\alpha_{\mathrm{s}}}-\frac{\boldsymbol{b}_{\mathrm{s}}}{\beta_{\mathrm{s}}}\Big)$ with $\boldsymbol{a}_{\mathrm{s}}$, $\alpha_{\mathrm{s}}$, $\boldsymbol{b}_{\mathrm{s}}$ and $\beta_{\mathrm{s}}$ as in (65). |
| **MTPV** | $\boldsymbol{i}_{\mathrm{s,ref}}^{k,\mathrm{MTPV}}(\gamma^\star) \overset{(69)}{:=} -2\big[\boldsymbol{D}-\gamma^\star\boldsymbol{J}\big]^{-1}\boldsymbol{d}$ | see Appendix D (case (i) for $\mathbb{MTPV}(\omega_{\mathrm{k}})\cap\partial\mathbb{V}(\omega_{\mathrm{k}},\hat{u}_{\max})$)<br>where $\boldsymbol{A}=\boldsymbol{M}_{\mathrm{V}}(\omega_{\mathrm{k}})$, $\boldsymbol{a}=\boldsymbol{m}_{\mathrm{V}}(\omega_{\mathrm{k}})$, $\alpha=\mu_{\mathrm{V}}(\omega_{\mathrm{k}})$,<br>$\boldsymbol{B}=\boldsymbol{V}(\omega_{\mathrm{k}})$, $\boldsymbol{b}=\boldsymbol{v}(\omega_{\mathrm{k}})$, $\beta=\nu(\omega_{\mathrm{k}},\hat{u}_{\max})$,<br>$\boldsymbol{D}=\big(\frac{\boldsymbol{A}}{\alpha}-\frac{\boldsymbol{B}}{\beta}\big)$ and $\boldsymbol{d}=\big(\frac{\boldsymbol{a}}{\alpha}-\frac{\boldsymbol{b}}{\beta}\big)$ |
| for RSM | $\boldsymbol{i}_{\mathrm{s,ref}}^{k,\mathrm{MTPV}}(\gamma^\star) \overset{(69)}{:=} -2\big[\boldsymbol{D}-\gamma^\star\boldsymbol{J}\big]^{-1}\boldsymbol{d}+\boldsymbol{x}_{\mathrm{s}}$ | see Appendix D (case (iii) for $\mathbb{MTPV}(\omega_{\mathrm{k}})\cap\partial\mathbb{V}(\omega_{\mathrm{k}},\hat{u}_{\max})$)<br>where $\boldsymbol{A}=\boldsymbol{M}_{\mathrm{V}}(\omega_{\mathrm{k}})$, $\boldsymbol{a}=\boldsymbol{m}_{\mathrm{V}}(\omega_{\mathrm{k}})=\boldsymbol{0}_2$, $\alpha=\mu_{\mathrm{V}}(\omega_{\mathrm{k}})=0$,<br>$\boldsymbol{x}_{\mathrm{s}}\neq\boldsymbol{0}_2$ (arbitrary), $\boldsymbol{B}=\boldsymbol{V}(\omega_{\mathrm{k}})$, $\boldsymbol{b}=\boldsymbol{v}(\omega_{\mathrm{k}})=\boldsymbol{0}_2$, $\beta=\nu(\omega_{\mathrm{k}},\hat{u}_{\max})$,<br>$\boldsymbol{D}:=\Big(\frac{\boldsymbol{A}}{\alpha_{\mathrm{s}}}-\frac{\boldsymbol{B}}{\beta_{\mathrm{s}}}\Big)$, $\boldsymbol{d}:=\Big(\frac{\boldsymbol{a}_{\mathrm{s}}}{\alpha_{\mathrm{s}}}-\frac{\boldsymbol{b}_{\mathrm{s}}}{\beta_{\mathrm{s}}}\Big)$ with $\boldsymbol{a}_{\mathrm{s}}$, $\alpha_{\mathrm{s}}$, $\boldsymbol{b}_{\mathrm{s}}$ and $\beta_{\mathrm{s}}$ as in (65). |

Table I: *Analytical solutions of the (optimal) current reference vectors for all operation strategies.*

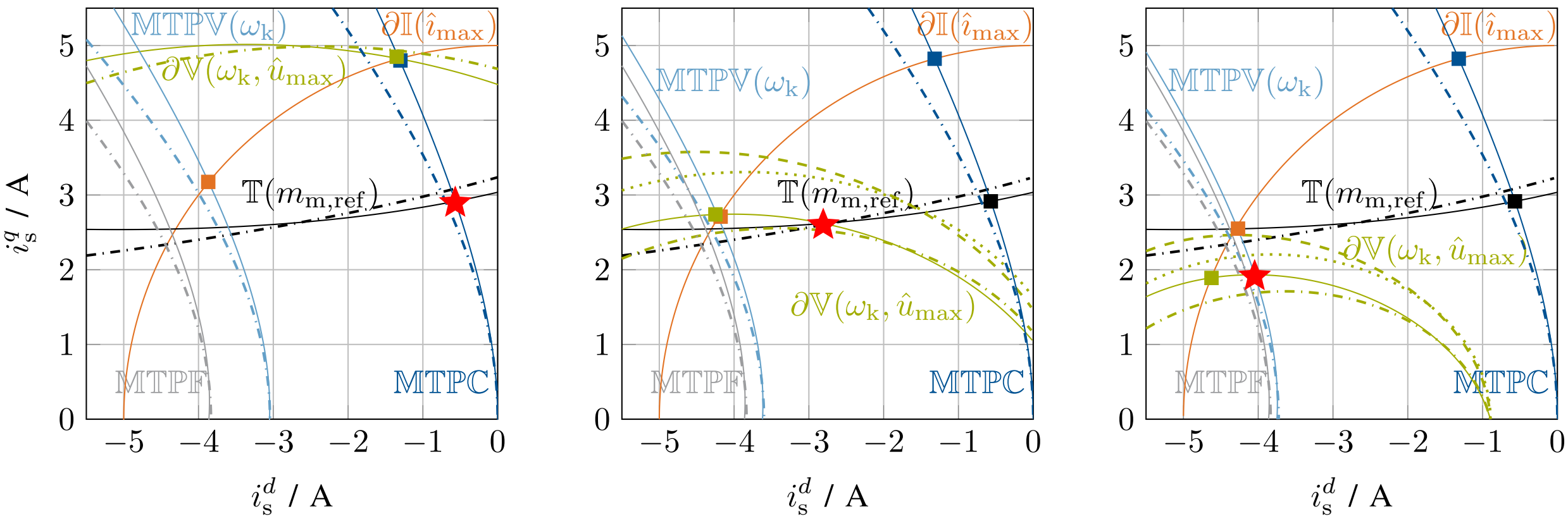

(a) *Maximum-Torque-per-Current (MTPC), i.e.,* $\mathbb{MTPC} \cap \mathbb{T}(m_{m,ref})$*, for* $\omega_k = \omega_{k,nom} \leq \omega_{k,nom}$.

(b) *Field weakening (FW), i.e.,* $\mathbb{T}(m_{m,ref}) \cap \partial\mathbb{V}(\omega_k, \hat{u}_{max})$*, for* $\omega_{k,nom} < \omega_k = 2\omega_{k,nom} < \omega_{k,cut\text{-}in}^{MTPV}$.

(c) *Maximum-Torque-per-Voltage (MTPV), i.e.,* $\mathbb{MTPV}(\omega_k) \cap \partial\mathbb{V}(\omega_k, \hat{u}_{max})$*, for* $\omega_k = 3\omega_{k,nom} \geq \omega_{k,cut\text{-}in}^{MTPV}$ *(*$m_{m,ref}$ *is* not *feasible anymore).*

Figure 4: *Illustration of the impact of* neglecting *stator resistance (dashed line:* $R_s = 0$*), mutual inductance (dash-dotted:* $L_m = 0$*) or both (dotted:* $R_s = L_m = 0$*)) on the different feedforward torque control strategies from Fig. 3: The three plots show voltage ellipse* —— $\partial\mathbb{V}(\omega_k, \hat{u}_{max})$, - - - $\partial\mathbb{V}(\omega_k, \hat{u}_{max}; R_s = 0)$, -·-· $\partial\mathbb{V}(\omega_k, \hat{u}_{max}; L_m = 0)$, ····· $\partial\mathbb{V}(\omega_k, \hat{u}_{max}; R_s = L_m = 0)$*, maximum current circle* —— $\partial\mathbb{I}(\hat{\imath}_{max})$*, MTPC hyperbola* —— $\mathbb{MTPC}$, -·-· $\mathbb{MTPC}(L_m = 0)$*, torque hyperbola* —— $\mathbb{T}(m_{m,ref})$, -·-· $\mathbb{T}(m_{m,ref}; L_m = 0)$*, MTPV hyperbola* —— $\mathbb{MTPV}(\omega_k)$, -·-· $\mathbb{MTPV}(\omega_k; L_m = 0)$*, MTPF hyperbola* —— $\mathbb{MTPF}$, -·-· $\mathbb{MTPF}(L_m = 0)$ *and optimal operation point* ★ *, respectively.*

### G. Graphical illustration of the operation strategies MPTC, FW and MTPV

In Fig. 3, for a 400 W synchronous machine with the following parameters

$$R_s = 20\,\Omega,\; L_s^d = 6 \cdot 10^{-2}\,\mathrm{H},\; L_s^q = 8 \cdot 10^{-2}\,\mathrm{H}, L_m = 0.5 \cdot 10^{-3}\,\mathrm{H}, \boldsymbol{\psi}_{pm}^k = (\psi_{pm}, 0)^\top = (0.23\,\mathrm{Wb}, 0)^\top \text{ and } n_p = 3, \quad (36)$$

three different optimal feedforward torque control strategies are illustrated for positive reference torque $m_{m,ref} = 3.35\,\mathrm{N\,m}$, voltage limit $\hat{u}_{max} = 600\,\mathrm{V}$ and current limit $\hat{\imath}_{max} = 5\,\mathrm{A}$. The illustrated optimal operation strategies are MTPC in Fig. 3a, FW in Fig. 3b, and MTPV in Fig. 3c. The respective optimal operation point, with its optimal reference current vector $\boldsymbol{i}_{s,ref}^k = (i_{s,ref}^d, i_{s,ref}^q)^\top$, is marked by ★ and corresponds to the intersection of (a) $\mathbb{MTPC} \cap \mathbb{T}(m_{m,ref})$ for MTPC in Fig. 3a, (b) $\partial\mathbb{V}(\omega_k, \hat{u}_{max}) \cap \mathbb{T}(m_{m,ref})$ for FW in Fig. 3b, and (c) $\mathbb{MTPV} \cap \partial\mathbb{V}(\omega_k, \hat{u}_{max})$ for MTPV in Fig. 3c.

For increasing electrical angular velocities $\omega_k \in \{1, 2, 3\}\omega_{k,nom}$ (where $\omega_{k,nom}$ is the nominal angular velocity), the MTPV hyperbola is approaching the MTPF hyperbola (recall Remark IV.6) and the voltage ellipse is shrinking; whereas the current circle, MTPC hyperbola, torque hyperbola and MTPF hyperbola are independent of the angular velocity and, hence, do *not* change in the three plots. The blue square represents the intersection of the MTPC hyperbola with the current circle and gives the nominal current vector $\boldsymbol{i}_{s,nom}^k$ producing the nominal torque $m_{m,nom}(\boldsymbol{i}_{s,nom}^k)$. The orange square highlights the intersection of current circle and torque hyperbola (in the $2^{nd}$ quadrant) and represents the maximally feasible current and maximally feasible torque for higher angular velocities.

Fig. 4 illustrates the impact of neglecting (i) stator resistance (i.e. $R_s = 0$: dashed line), (ii) mutual inductance (i.e. $L_m = 0$: dash-dotted line) or (iii) both (i.e. $R_s = L_m = 0$: dotted line) on the shape of MTPC, MTPV, MTPF and torque hyperbolas and the voltage ellipse. The feedforward torque control strategies and optimal operation points (marked by ★ ) are identical to those shown in Fig. 3. It is easy to see that neglecting stator resistance, mutual inductance or both will lead to completely different (and wrong) intersection points and, hence, *not* optimal operation points with *reduced efficiency*. For example, the impact of neglecting stator resistance, mutual inductance or both on the shape, size and orientation of the voltage ellipse is obvious. Concluding, for optimal operation of a synchronous machine, both parameters must *not* be neglected.

In Fig. 3 and 4, the intersection points of (a) current circle and MTPC hyperbola, (b) current circle and MTPV hyperbola, (c) current circle and voltage ellipse, and (d) torque and MTPC hyperbola are highlighted by the following colored squares (a) ■, (b) ■, (c) ■, and (d) ■, respectively.

### Acknowledgement

This project has received funding from the European Union's Horizon 2020 research and innovation programme under the Marie Sklodowska-Curie grant agreement No. 642682 and from the Bavarian Ministry for Eduction, Culture, Science, and Art.

## APPENDIX

In this appendix, all necessary mathematical derivations are presented for two general quadrics given by

$$Q_A(\boldsymbol{x}) := \boldsymbol{x}^\top \boldsymbol{A}\boldsymbol{x} + 2\boldsymbol{a}^\top \boldsymbol{x} + \alpha \qquad \text{and} \qquad Q_B(\boldsymbol{x}) := \boldsymbol{x}^\top \boldsymbol{B}\boldsymbol{x} + 2\boldsymbol{b}^\top \boldsymbol{x} + \beta, \tag{37}$$

where

$$\boldsymbol{A} = \boldsymbol{A}^\top := \begin{bmatrix} a_{11} & a_{12} \\ a_{12} & a_{22} \end{bmatrix} \in \mathbb{R}^{2\times 2},\ \boldsymbol{a} := \begin{pmatrix} a_1 \\ a_2 \end{pmatrix} \in \mathbb{R}^2,\ \alpha \in \mathbb{R},\ \boldsymbol{B} = \boldsymbol{B}^\top := \begin{bmatrix} b_{11} & b_{12} \\ b_{12} & b_{22} \end{bmatrix} \in \mathbb{R}^{2\times 2},\ \boldsymbol{b} := \begin{pmatrix} b_1 \\ b_2 \end{pmatrix} \in \mathbb{R}^2 \text{ and } \beta \in \mathbb{R}. \tag{38}$$

Main goal is to explain in detail how analytical solutions to the MTPC, MTPV and MTPF optimization problems can be obtained and how analytical solutions for the intersection points of two general quadrics can be found. As it will be shown, all problems can be solved by finding the roots of a fourth-order *quartic polynomial* for which (luckily) still an analytical solutions exists.

### A. Formulation of the optimization problem with equality constraints

All optimization problems like MTPC, MTPV and MTPF can be formulated in a general framework as optimization problems with equality constraint by invoking the quadrics $Q_A(\boldsymbol{i}_\mathrm{s}^k)$ and $Q_B(\boldsymbol{i}_\mathrm{s}^k)$ in (37) as follows

$$\boldsymbol{i}_{\mathrm{s,ref}}^k := \arg\max_{\boldsymbol{i}_\mathrm{s}^k} -\Big(\underbrace{(\boldsymbol{i}_\mathrm{s}^k)^\top \boldsymbol{A}\boldsymbol{i}_\mathrm{s}^k + 2\boldsymbol{a}^\top \boldsymbol{i}_\mathrm{s}^k + \alpha}_{=:Q_A(\boldsymbol{i}_\mathrm{s}^k)}\Big) \qquad \text{s.t.} \qquad \underbrace{(\boldsymbol{i}_\mathrm{s}^k)^\top \boldsymbol{B}\boldsymbol{i}_\mathrm{s}^k + 2\boldsymbol{b}^\top \boldsymbol{i}_\mathrm{s}^k + \beta}_{=:Q_B(\boldsymbol{i}_\mathrm{s}^k)} = 0. \tag{39}$$

A first idea, based on quadrics, was presented in [42] for the MTPC strategy. The optimization problem (39) can be reformulated as *Lagrangian* (see [43, Cha. 5]) where the (possibly complex) Lagrangian multiplier $\lambda \in \mathbb{C}$ must be introduced for the equality constraint, i.e.

$$\mathfrak{L}(\boldsymbol{i}_\mathrm{s}^k, \lambda) := -Q_A(\boldsymbol{i}_\mathrm{s}^k) + \lambda\, Q_B(\boldsymbol{i}_\mathrm{s}^k) \overset{(39)}{=} -\big[(\boldsymbol{i}_\mathrm{s}^k)^\top \boldsymbol{A}\boldsymbol{i}_\mathrm{s}^k + 2\boldsymbol{a}^\top \boldsymbol{i}_\mathrm{s}^k + \alpha\big] + \lambda\ \big[(\boldsymbol{i}_\mathrm{s}^k)^\top \boldsymbol{B}\boldsymbol{i}_\mathrm{s}^k + 2\boldsymbol{b}^\top \boldsymbol{i}_\mathrm{s}^k + \beta\big]. \tag{40}$$

For the three different optimization problems MTPC, MTPF and MTPV, the matrices $\boldsymbol{A}$, $\boldsymbol{B}$, the vectors $\boldsymbol{a}$, $\boldsymbol{b}$ and the scalars $\alpha$, $\beta$ must be chosen properly according to the considered optimization problem as specified in the following:

- MTPC: $\boldsymbol{A} = \boldsymbol{I}_2$, $\boldsymbol{a} = (0,\, 0)^\top$, $\alpha = 0$ and $\boldsymbol{B} = \boldsymbol{T}$, $\boldsymbol{b} = \boldsymbol{t}$ and $\beta = \tau(m_{\mathrm{m,ref}})$,
- MTPV: $\boldsymbol{A} = \boldsymbol{V}(\omega_\mathrm{k})$, $\boldsymbol{a} = \boldsymbol{v}(\omega_\mathrm{k})$, $\alpha = \nu(\omega_\mathrm{k}, \hat{u}_{\max})$ and $\boldsymbol{B} = \boldsymbol{T}$, $\boldsymbol{b} = \boldsymbol{t}$ and $\beta = \tau(m_{\mathrm{m,ref}})$, and
- MTPF: $\boldsymbol{A} = \boldsymbol{F}$, $\boldsymbol{a} = \boldsymbol{f}$, $\alpha = \phi$ and $\boldsymbol{B} = \boldsymbol{T}$, $\boldsymbol{b} = \boldsymbol{t}$ and $\beta = \tau(m_{\mathrm{m,ref}})$,

with $\boldsymbol{T}$, $\boldsymbol{t}$ and $\tau(m_{\mathrm{m,ref}})$ as in (9), $\boldsymbol{V}(\omega_\mathrm{k})$, $\boldsymbol{v}(\omega_\mathrm{k})$ and $\nu(\omega_\mathrm{k}, \hat{u}_{\max})$ as in (15) and $\boldsymbol{F}$, $\boldsymbol{f}$ and $\phi$ as in (21). To obtain the optimal (maximizing) reference current vector as in (39), the following necessary and sufficient conditions must be evaluated.

*1) Necessary condition for a maximum:* To find a maximum, the following *necessary* condition must be satisfied: The gradient[4] of the Lagrangian must be equal to the zero vector, i.e.

$$\boldsymbol{g}_{\mathfrak{L}}(\boldsymbol{i}_\mathrm{s}^k, \lambda) := \left(\frac{\mathrm{d}\mathfrak{L}(\boldsymbol{i}_\mathrm{s}^k, \lambda)}{\mathrm{d}(\boldsymbol{i}_\mathrm{s}^k, \lambda)}\right)^\top = \begin{pmatrix} \left(\dfrac{\mathrm{d}\mathfrak{L}(\boldsymbol{i}_\mathrm{s}^k, \lambda)}{\mathrm{d}\boldsymbol{i}_\mathrm{s}^k}\right)^\top \\ \dfrac{\mathrm{d}\mathfrak{L}(\boldsymbol{i}_\mathrm{s}^k, \lambda)}{\mathrm{d}\lambda} \end{pmatrix} \overset{!}{=} \boldsymbol{0}_3 \quad \overset{(39)}{\Longrightarrow} \quad \begin{pmatrix} -2\boldsymbol{A}\boldsymbol{i}_\mathrm{s}^k - 2\boldsymbol{a} + \lambda(2\boldsymbol{B}\boldsymbol{i}_\mathrm{s}^k + 2\boldsymbol{b}) \\ (\boldsymbol{i}_\mathrm{s}^k)^\top \boldsymbol{B}\boldsymbol{i}_\mathrm{s}^k + 2\boldsymbol{b}^\top \boldsymbol{i}_\mathrm{s}^k + \beta \end{pmatrix} \overset{!}{=} \boldsymbol{0}_3. \tag{41}$$

By defining

$$\boldsymbol{M}(\lambda) := \lambda\boldsymbol{B} - \boldsymbol{A} = \begin{bmatrix} \lambda\, b_{11} - a_{11} & \lambda\, b_{12} - a_{12} \\ \lambda\, b_{12} - a_{12} & \lambda\, b_{22} - a_{22} \end{bmatrix} \qquad \text{and} \qquad \boldsymbol{m}(\lambda) := \lambda\,\boldsymbol{b} - \boldsymbol{a} = \begin{pmatrix} \lambda\, b_1 - a_1 \\ \lambda\, b_2 - a_2 \end{pmatrix}, \tag{42}$$

one may rewrite the first two rows in (41) in the compact form

$$2\big[\lambda\boldsymbol{B} - \boldsymbol{A}\big]\boldsymbol{i}_\mathrm{s}^k + \big(\lambda\,\boldsymbol{b} - \boldsymbol{a}\big) \overset{(42)}{=} 2\boldsymbol{M}(\lambda)\boldsymbol{i}_\mathrm{s}^k + 2\boldsymbol{m}(\lambda) = \boldsymbol{0}_2 \tag{43}$$

and solving for $\boldsymbol{i}_\mathrm{s}^k = \boldsymbol{i}_\mathrm{s}^{k,\star}$ yields

$$\boldsymbol{i}_\mathrm{s}^{k,\star}(\lambda) = -\big[\lambda\boldsymbol{B} - \boldsymbol{A}\big]^{-1}\big(\lambda\,\boldsymbol{b} - \boldsymbol{a}\big) \overset{(42)}{=} -\boldsymbol{M}(\lambda)^{-1}\boldsymbol{m}(\lambda) \tag{44}$$

where

$$\boldsymbol{M}(\lambda)^{-1} = \big[\lambda\boldsymbol{B} - \boldsymbol{A}\big]^{-1} = \frac{1}{\det \boldsymbol{M}(\lambda)} \begin{bmatrix} \lambda\, b_{22} - a_{22} & -\lambda\, b_{12} + a_{12} \\ -\lambda\, b_{12} + a_{12} & \lambda\, b_{11} - a_{11} \end{bmatrix} = \big(\big[\lambda\boldsymbol{B} - \boldsymbol{A}\big]^{-1}\big)^\top = \big(\boldsymbol{M}(\lambda)^{-1}\big)^\top$$

[4]Note that, for some vectors $\boldsymbol{x}, \boldsymbol{c} \in \mathbb{R}^n$ and a symmetric matrix $\boldsymbol{M} = \boldsymbol{M}^\top \in \mathbb{R}^{n\times n}$, the following hold $(\frac{\mathrm{d}\boldsymbol{c}^\top\boldsymbol{x}}{\mathrm{d}\boldsymbol{x}})^\top = (\frac{\mathrm{d}\boldsymbol{c}^\top\boldsymbol{x}}{\mathrm{d}x_1}, \ldots, \frac{\mathrm{d}\boldsymbol{c}^\top\boldsymbol{x}}{\mathrm{d}x_n})^\top = (c_1, \ldots, c_n)^\top = \boldsymbol{c}$ and $(\frac{\mathrm{d}\boldsymbol{x}^\top\boldsymbol{M}\boldsymbol{x}}{\mathrm{d}\boldsymbol{x}})^\top = (\frac{\mathrm{d}\boldsymbol{x}^\top\boldsymbol{M}\boldsymbol{x}}{\mathrm{d}x_1}, \ldots, \frac{\mathrm{d}\boldsymbol{x}^\top\boldsymbol{M}\boldsymbol{x}}{\mathrm{d}x_n})^\top = (\boldsymbol{M} + \boldsymbol{M}^\top)\boldsymbol{x} = 2\boldsymbol{M}\boldsymbol{x}$ (see [44, Proposition 10.7.1 i)]).

and

$$\det \boldsymbol{M}(\lambda) = (\lambda\, b_{11} - a_{11})(\lambda\, b_{22} - a_{22}) - (\lambda\, b_{12} - a_{12})^2 = \big(\det \boldsymbol{B}\big)\lambda^2 + \big(\det\big(\boldsymbol{B} - \boldsymbol{A}\big) - \det \boldsymbol{A} - \det \boldsymbol{B}\big)\lambda + \det \boldsymbol{A}.$$

Inserting $\boldsymbol{i}_{\mathrm{s}}^k = \boldsymbol{i}_{\mathrm{s}}^{k,\star}(\lambda)$ as in (44) into the constraint quadric $Q_B(\boldsymbol{i}_{\mathrm{s}}^k)$ as in (39) gives a quartic polynomial in $\lambda$ as follows

$$\left.\begin{array}{rcl} \boldsymbol{m}(\lambda)^\top \boldsymbol{M}(\lambda)^{-1}\boldsymbol{B}\,\boldsymbol{M}(\lambda)^{-1}\boldsymbol{m}(\lambda) - 2\boldsymbol{b}^\top \boldsymbol{M}(\lambda)^{-1}\boldsymbol{m}(\lambda) + \beta & = & 0 \quad \Big| \quad \cdot \det\big(\boldsymbol{M}(\lambda)\big)^2 \\ \Longrightarrow \qquad \chi_4(\lambda) := c_4\lambda^4 \,+\, c_3\lambda^3 \,+\, c_2\lambda^2 \,+\, c_1\lambda \,+\, c_0 = 0 & & \end{array}\right\} \tag{45}$$

with *real* coefficients

$$\left.\begin{array}{rcl}
c_4 & := & -\left(b_{11}b_{22} - b_{12}^2\right)\left(b_{22}b_1^2 - 2b_1b_2b_{12} + b_{11}b_2^2 + \beta b_{12}^2 - b_{11}b_{22}\beta\right) \\
c_3 & := & 2(a_{11}b_{22} - 2a_{12}b_{12} + a_{22}b_{11})(b_{22}b_1^2 - 2b_1b_2b_{12} + b_{11}b_2^2 + \beta b_{12}^2 - b_{11}b_{22}\beta) \\
c_2 & := & a_1^2b_{11}b_{22}^2 - a_1^2b_{12}^2b_{22} - 2a_1a_2b_{11}b_{12}b_{22} + 2a_1a_2b_{12}^3 - 2a_1a_{11}b_1b_{22}^2 \\
& & +2a_1a_{11}b_2b_{12}b_{22} + 4a_1a_{12}b_1b_{12}b_{22} - 2a_1a_{12}b_2b_{11}b_{22} - 2a_1a_{12}b_2b_{12}^2 \\
& & -2a_1a_{22}b_1b_{12}^2 + 2a_1a_{22}b_2b_{11}b_{12} + a_2^2b_{11}^2b_{22} - a_2^2b_{11}b_{12}^2 + 2a_2a_{11}b_1b_{12}b_{22} \\
& & -2a_2a_{11}b_2b_{12}^2 - 2a_2a_{12}b_1b_{11}b_{22} - 2a_2a_{12}b_1b_{12}^2 + 4a_2a_{12}b_2b_{11}b_{12} \\
& & +2a_2a_{22}b_1b_{11}b_{12} - 2a_2a_{22}b_2b_{11}^2 - a_{11}^2b_2^2b_{22} + \beta a_{11}^2b_{22}^2 + 2a_{11}a_{12}b_1b_2b_{22} \\
& & +2a_{11}a_{12}b_2^2b_{12} - 4\beta a_{11}a_{12}b_{12}b_{22} - 4a_{11}a_{22}b_1^2b_{22} + 6a_{11}a_{22}b_1b_2b_{12} \\
& & -4a_{11}a_{22}b_2^2b_{11} + 4\beta a_{11}a_{22}b_{11}b_{22} - 2\beta a_{11}a_{22}b_{12}^2 + 3a_{12}^2b_1^2b_{22} - 10a_{12}^2b_1b_2b_{12} \\
& & +3a_{12}^2b_2^2b_{11} - 2\beta a_{12}^2b_{11}b_{22} + 6\beta a_{12}^2b_{12}^2 + 2a_{12}a_{22}b_1^2b_{12} + 2a_{12}a_{22}b_1b_2b_{11} \\
& & -4\beta a_{12}a_{22}b_{11}b_{12} - a_{22}^2b_1^2b_{11} + \beta a_{22}^2b_{11}^2 \\
c_1 & := & 2a_1^2a_{22}b_{12}^2 - 2b_{11}b_{22}a_1^2a_{22} - 4a_1a_2a_{12}b_{12}^2 + 4b_{11}b_{22}a_1a_2a_{12} + 4b_{22}a_1a_{11}a_{22}b_1 \\
& & -4a_1a_{11}a_{22}b_2b_{12} - 4b_{22}a_1a_{12}^2b_1 + 4a_1a_{12}^2b_2b_{12} + 2a_2^2a_{11}b_{12}^2 - 2b_{11}b_{22}a_2^2a_{11} \\
& & -4a_2a_{11}a_{22}b_1b_{12} + 4b_{11}a_2a_{11}a_{22}b_2 + 4a_2a_{12}^2b_1b_{12} - 4b_{11}a_2a_{12}^2b_2 + 2a_{11}^2a_{22}b_2^2 \\
& & -2b_{22}\beta a_{11}^2a_{22} - 2a_{11}a_{12}^2b_2^2 + 2b_{22}\beta a_{11}a_{12}^2 - 4a_{11}a_{12}a_{22}b_1b_2 + 4\beta a_{11}a_{12}a_{22}b_{12} \\
& & +2a_{11}a_{22}^2b_1^2 - 2b_{11}\beta a_{11}a_{22}^2 + 4a_{12}^3b_1b_2 - 4\beta a_{12}^3b_{12} - 2a_{12}^2a_{22}b_1^2 + 2b_{11}\beta a_{12}^2a_{22} \\
c_0 & := & b_{22}a_1^2a_{12}^2 - 2b_{12}a_1^2a_{12}a_{22} + b_{11}a_1^2a_{22}^2 - 2b_{22}a_1a_2a_{11}a_{12} + 2b_{12}a_1a_2a_{11}a_{22} + 2b_{12}a_1a_2a_{12}^2 \\
& & -2b_{11}a_1a_2a_{12}a_{22} + 2b_2a_1a_{11}a_{12}a_{22} - 2b_1a_1a_{11}a_{22}^2 - 2b_2a_1a_{12}^3 + 2b_1a_1a_{12}^2a_{22} + b_{22}a_2^2a_{11}^2 \\
& & -2b_{12}a_2^2a_{11}a_{12} + b_{11}a_2^2a_{12}^2 - 2b_2a_2a_{11}^2a_{22} + 2b_2a_2a_{11}a_{12}^2 + 2b_1a_2a_{11}a_{12}a_{22} - 2b_1a_2a_{12}^3 \\
& & +\beta a_{11}^2a_{22}^2 - 2\beta a_{11}a_{12}^2a_{22} + \beta a_{12}^4.
\end{array}\right\} \tag{46}$$

*2) Sufficient condition for a maximum:* To obtain a *maximum* under an equality constraint, the Hessian of the Lagrangian $\mathfrak{L}(\boldsymbol{i}_{\mathrm{s}}^k, \lambda)$ must be *negative definite*, i.e.

$$\boldsymbol{H}_{\mathfrak{L}}(\boldsymbol{i}_{\mathrm{s}}^k, \lambda) := \tfrac{\mathrm{d}}{\mathrm{d}(\boldsymbol{i}_{\mathrm{s}}^k,\lambda)} \left(\tfrac{\mathrm{d}\mathfrak{L}(\boldsymbol{i}_{\mathrm{s}}^k,\lambda)}{\mathrm{d}(\boldsymbol{i}_{\mathrm{s}}^k,\lambda)}\right)^\top \overset{!}{<} 0. \tag{47}$$

The Hessian matrix is symmetric and given by

$$\boldsymbol{H}_{\mathfrak{L}}(\boldsymbol{i}_{\mathrm{s}}^k, \lambda) := \tfrac{\mathrm{d}}{\mathrm{d}(\boldsymbol{i}_{\mathrm{s}}^k,\lambda)} \left(\tfrac{\mathrm{d}\mathfrak{L}(\boldsymbol{i}_{\mathrm{s}}^k,\lambda)}{\mathrm{d}(\boldsymbol{i}_{\mathrm{s}}^k,\lambda)}\right)^\top = \begin{bmatrix} 2\boldsymbol{M}(\lambda) & 2\boldsymbol{B}\boldsymbol{i}_{\mathrm{s}}^k + 2\boldsymbol{b} \\ (2\boldsymbol{B}\boldsymbol{i}_{\mathrm{s}}^k + 2\boldsymbol{b})^\top & 0 \end{bmatrix} = \boldsymbol{H}_{\mathfrak{L}}(\boldsymbol{i}_{\mathrm{s}}^k, \lambda)^\top \in \mathbb{R}^{3\times 3}. \tag{48}$$

The Hessian $\boldsymbol{H}_{\mathfrak{L}}(\boldsymbol{i}_{\mathrm{s}}^k, \lambda)$ is negative definite if and only if all its leading principal minors[5] have alternating signs (see [44, Prop. 8.2.8] in combination with [44, Prop. 2.7.1][6]). More precisely, first and third leading principle minor of (48) must be *negative* whereas the second leading principle minor must be *positive*, i.e. $2m_{11}(\lambda^\star) \overset{(42)}{=} 2\big(a_{11} + \lambda^\star b_{11}\big) < 0$ (first leading principal minor), $\det\big(2\boldsymbol{M}(\lambda^\star)\big) = 2^2 \det \boldsymbol{M}(\lambda^\star) > 0$ (second leading principal minor) and $\det\big(\boldsymbol{H}_{\mathfrak{L}}(\boldsymbol{i}_{\mathrm{s}}^k, \lambda^\star)\big) < 0$ (third leading principal minor). Hence, the optimal (real) Lagrangian multiplier $\lambda^\star \in \mathbb{R}$ must satisfy

$$\left.\begin{array}{lllcrcl}
\text{(i)} & \lambda^\star & < \dfrac{a_{11}}{b_{11}} & \Longrightarrow & 2\big(\lambda^\star b_{11} - a_{11}\big) & < & 0 \\
\text{(ii)} & \lambda^\star & > \dfrac{\big(\det\big(\boldsymbol{B}-\boldsymbol{A}\big) - \det \boldsymbol{A} - \det \boldsymbol{B}\big)}{-\operatorname{sign}(\det \boldsymbol{B})\det \boldsymbol{B}}\left(1 \pm \sqrt{1 - \frac{4\det \boldsymbol{A}\det \boldsymbol{B}}{\big(\det\big(\boldsymbol{B}-\boldsymbol{A}\big) - \det \boldsymbol{A} - \det \boldsymbol{B}\big)^2}}\right) & \Longrightarrow & \det \boldsymbol{M}(\lambda^\star) & > & 0.
\end{array}\right\} \tag{49}$$

Clearly, the conditions (i) and (ii) in (49) must be satisfied simultaneously and, therefore, imply negative definiteness of $\boldsymbol{M}(\lambda^\star) = \lambda^\star \boldsymbol{B} - \boldsymbol{A}$. Moreover, by defining

$$\boldsymbol{C}(\boldsymbol{i}_{\mathrm{s}}^k, \lambda^\star) := \begin{bmatrix} \boldsymbol{I}_2, & \boldsymbol{M}(\lambda^\star)^{-1}(\boldsymbol{B}\boldsymbol{i}_{\mathrm{s}}^k + \boldsymbol{b}) \\ \boldsymbol{0}_2^\top, & 1 \end{bmatrix} \in \mathbb{R}^{3\times 3} \qquad \text{with} \qquad \det \boldsymbol{C}(\boldsymbol{i}_{\mathrm{s}}^k, \lambda^\star) = \det \boldsymbol{C}(\boldsymbol{i}_{\mathrm{s}}^k, \lambda^\star)^\top = 1$$

[5]The $i$-th leading principle minor is the determinant of the $(i, i)$-north-western submatrix of the matrix [44, Prop. 8.2.7].

[6]I.e., for all $\gamma \in \mathbb{R}$ and $\boldsymbol{A} \in \mathbb{R}^{n\times n}$, the following holds $\det(\gamma \boldsymbol{A}) = (\gamma)^n \det(\boldsymbol{A})$. Hence, application of the Sylvester's criterion to negative definite matrices yields alternating signs of the leading principle minors.

and by invoking [44, Fact 2.16.2], the Hessian matrix

$$\boldsymbol{H}_{\mathfrak{L}}(\boldsymbol{i}_{\rm s}^k,\lambda^\star) = 2\boldsymbol{C}(\boldsymbol{i}_{\rm s}^k,\lambda^\star)^\top \begin{bmatrix} \boldsymbol{M}(\lambda^\star), & \boldsymbol{0}_2 \\ \boldsymbol{0}_2^\top, & -\big((\boldsymbol{i}_{\rm s}^k)^\top \boldsymbol{B}^\top + \boldsymbol{b}^\top\big)\boldsymbol{M}(\lambda^\star)^{-1}\big(\boldsymbol{B}\boldsymbol{i}_{\rm s}^k + \boldsymbol{b}\big) \end{bmatrix} \boldsymbol{C}(\boldsymbol{i}_{\rm s}^k,\lambda^\star)$$

can be written as product of three matrices. Hence,

$$\det\big[\boldsymbol{H}_{\mathfrak{L}}(\boldsymbol{i}_{\rm s}^k,\lambda^\star)\big] = -2\underbrace{\big((\boldsymbol{i}_{\rm s}^k)^\top \boldsymbol{B}^\top + \boldsymbol{b}^\top\big)\boldsymbol{M}(\lambda^\star)^{-1}\big(\boldsymbol{B}\boldsymbol{i}_{\rm s}^k + \boldsymbol{b}\big)}_{=:\,\gamma\in\mathbb{R}} \cdot \det \boldsymbol{M}(\lambda^\star),$$

which, with $\det \boldsymbol{C}(\boldsymbol{i}_{\rm s}^k,\lambda^\star) = \det \boldsymbol{C}(\boldsymbol{i}_{\rm s}^k,\lambda^\star)^\top = 1$ and negative definiteness of $\boldsymbol{M}(\lambda^\star) = \big[\boldsymbol{A} + \lambda^\star \boldsymbol{B}\big] \overset{(49)}{<} 0$, implies that $\gamma = \|\boldsymbol{B}\boldsymbol{i}_{\rm s}^k + \boldsymbol{b}\|^2_{\boldsymbol{M}(\lambda^\star)} < 0$ (a weighted norm with negative definite $\boldsymbol{M}(\lambda^\star) < 0$) is negative for all *non-zero* vectors

$$\begin{aligned} \boldsymbol{B}\boldsymbol{i}_{\rm s}^{k,\star}(\lambda^\star) + \boldsymbol{b} \quad &\overset{(44)}{=} \quad -\lambda^\star \boldsymbol{B}\big[\boldsymbol{A} + \lambda^\star \boldsymbol{B}\big]^{-1}\boldsymbol{b} + \boldsymbol{b} = -\big(-\boldsymbol{I}_2 + \lambda^\star \boldsymbol{B}\big[\boldsymbol{A} + \lambda^\star \boldsymbol{B}\big]^{-1}\big)\boldsymbol{b} \\ &\overset{[44,\ \text{Cor. 2.8.10}]}{=} \quad \big[\boldsymbol{A} + \lambda^\star \boldsymbol{B}\big]^{-1}\boldsymbol{b} \overset{(49)}{\neq} \boldsymbol{0}_2 \quad \text{for all} \quad \lambda^\star \text{ as in (49).} \end{aligned}$$

Concluding, for the optimal $\lambda^\star$, the third leading principle minor is (always) negative, i.e. $\det\big[\boldsymbol{H}_{\mathfrak{L}}(\boldsymbol{i}_{\rm s}^k,\lambda^\star)\big] < 0$. By checking definiteness of (47) for $\lambda^\star \in \{\lambda_1^\star,\ldots,\lambda_4^\star\}$ where $\boldsymbol{M}(\lambda^\star) > 0$, the analytical solution for the optimal reference current vector is finally given by

$$\boxed{\boldsymbol{i}_{\rm s,ref}^k \overset{(39)}{:=} \boldsymbol{i}_{\rm s}^{k,\star}(\lambda^\star) \overset{(44)}{=} -\boldsymbol{M}(\lambda^\star)^{-1}\boldsymbol{m}(\lambda^\star) = -\big[\lambda^\star \boldsymbol{B} - \boldsymbol{A}\big]^{-1}\big(\lambda^\star \boldsymbol{b} - \boldsymbol{a}\big).} \tag{50}$$

**Remark A.1** (The case $\boldsymbol{m}(\lambda) = \boldsymbol{0}_2$ for all $\lambda \in \mathbb{C}$)**.** *Note that the optimal reference current vector* (50) *only gives a non-trivial solution if* $\boldsymbol{m}(\lambda) = \lambda\,\boldsymbol{b} - \boldsymbol{a} \neq 0$. *This is not true for RSMs, where* $\psi_{\rm pm} = 0$ *and, hence,* $\boldsymbol{t} \overset{(9)}{=} \boldsymbol{0}_2$, $\boldsymbol{v}(\omega_{\rm k}) \overset{(15)}{=} \boldsymbol{0}_2$ *and* $\boldsymbol{f} \overset{(21)}{=} \boldsymbol{0}_2$. *To solve these optimization problems,* (39) *must be re-formulated by shifting/translating the quadrics* $Q_A(\boldsymbol{i}_{\rm s}^k)$ *and* $Q_B(\boldsymbol{i}_{\rm s}^k)$ *by some non-zero but constant* $\boldsymbol{x}_{\rm s} \in \mathbb{R}^2$. *The shifted quadrics* $Q_A(\bar{\boldsymbol{i}}_{\rm s}^k + \boldsymbol{x}_{\rm s})$ *and* $Q_B(\bar{\boldsymbol{i}}_{\rm s}^k + \boldsymbol{x}_{\rm s})$ *are obtained by inserting* $\boldsymbol{i}_{\rm s}^k = \bar{\boldsymbol{i}}_{\rm s}^k + \boldsymbol{x}_{\rm s}$ *into* (39) *(for more details see Appendix D).*

**Remark A.2** (Normalization)**.** *For a numerical implementation, a normalization of* (8) *might be beneficial yielding a less ill-conditioned optimization problem (the coefficients of the fourth-order polynomials heavily differ in magnitude). The normalized version of* (8) *is*

$$\bar{\boldsymbol{u}}_{\rm s}^k = \bar{R}_{\rm s}\bar{\boldsymbol{i}}_{\rm s}^k + \bar{\omega}_{\rm k}\boldsymbol{J}\bar{\boldsymbol{L}}_{\rm s}^k\bar{\boldsymbol{i}}_{\rm s}^k + \bar{\omega}_{\rm k}\boldsymbol{J}\bar{\boldsymbol{\psi}}_{\rm pm}^k \tag{51}$$

*where the normalized (unitless) quantities and parameters are defined by*

$$\bar{\boldsymbol{u}}_{\rm s}^k := \frac{\boldsymbol{u}_{\rm s}^k}{\hat{u}_{\max}},\quad \bar{R}_{\rm s} := \frac{R_{\rm s}}{\hat{u}_{\max}/\hat{\imath}_{\max}},\quad \bar{\boldsymbol{i}}_{\rm s}^k := \frac{\boldsymbol{i}_{\rm s}^k}{\hat{\imath}_{\max}},\quad \bar{\omega}_{\rm k} := \frac{\omega_{\rm k}}{\omega_{\rm k,nom}},\quad \bar{\boldsymbol{L}}_{\rm s}^k := \frac{\omega_{\rm k,nom}}{\hat{u}_{\max}/\hat{\imath}_{\max}}\boldsymbol{L}_{\rm s}^k \ \text{ and } \ \bar{\boldsymbol{\psi}}_{\rm pm}^k := \frac{\omega_{\rm k,nom}}{\hat{u}_{\max}}\boldsymbol{\psi}_{\rm pm}^k.$$

*The normalized (unitless) machine torque and its reference are as follows*

$$\bar{m}_{\rm m} := \frac{\omega_{\rm k,nom}}{\hat{u}_{\max}\hat{\imath}_{\max}} m_{\rm m} \qquad \text{and} \qquad \bar{m}_{\rm m,ref} := \frac{\omega_{\rm k,nom}}{\hat{u}_{\max}\hat{\imath}_{\max}} m_{\rm m,ref}.$$

*3) Analytical computation of the roots of a quartic polynomial:* The discriminant of the quartic polynomial $\chi_4(\lambda)$ as in (45) can be computed as follows

$$\begin{aligned} \Delta := 256c_4^3c_0^3 - 192c_4^2c_3c_1c_0^2 - 128c_4^2c_2^2c_0^2 + 144c_4^2c_2c_1^2c_0 - 27c_4^2c_1^4 \\ + 144c_4c_3^2c_2c_0^2 - 6c_4c_3^2c_1^2c_0 - 80c_4c_3c_2^2c_1c_0 + 18c_4c_3c_2c_1^3 + 16c_4c_2^4c_0 \\ - 4c_4c_2^3c_1^2 - 27c_3^4c_0^2 + 18c_3^3c_2c_1c_0 - 4c_3^3c_1^3 - 4c_3^2c_2^3c_0 + c_3^2c_2^2c_1^2. \end{aligned} \tag{52}$$

For $\Delta < 0$, the quartic polynomial (45) has two real and two complex roots (all distinct); for $\Delta > 0$, (45) has four real or four complex roots (all distinct) and, for $\Delta = 0$, (45) has at least two equal roots (for more details see [41]). Euler's solution will be presented (for details see [45]) which is based on the *depressed* (and monic) quartic polynomial given by

$$\chi_{4,\rm dep}(y) := y^4 + py^2 + qy + r = 0 \tag{53}$$

with real coefficients

$$p := \tfrac{1}{c_4^2}\Big(c_2\,c_4 - \tfrac{3\,c_3^2}{8}\Big),\quad q := \tfrac{1}{c_4^3}\Big(\tfrac{c_3^3}{8} - \tfrac{c_2\,c_3\,c_4}{2} + c_1\,c_4^2\Big) \quad \text{and} \quad r := \tfrac{1}{c_4^4}\Big(-\tfrac{3c_3^4}{256} + c_4^3c_0 - \tfrac{c_4^2c_3c_2}{4} + \tfrac{c_4c_3^2c_2}{16}\Big). \tag{54}$$

The depressed quartic polynomial (53) is obtained by inserting $\lambda := y - \frac{c_3}{4c_4}$ into (45). To compute the roots of the depressed quartic, one needs to find the three roots $z_1^\star$, $z_2^\star$ and $z_3^\star$ of Euler's resolvent cubic polynomial given by

$$\chi_{3,\rm res}(z) := z^3 + 2p\,z^2 + (p^2 - 4r)\,z - q^2 = 0, \tag{55}$$

where $p$, $q$ and $r$ are as in (54). In Appendix A4, the analytical solution to compute the three roots $z_1^\star$, $z_2^\star$ and $z_3^\star$ of the resolvent cubic (55) is presented. Finally, for known resolvent roots $z_1^\star$, $z_2^\star$ and $z_3^\star$, the four roots $\lambda_1^\star$, $\lambda_2^\star$, $\lambda_3^\star$ and $\lambda_4^\star$ of the quartic polynomial (45) are given by

$$\left.\begin{aligned}
\lambda_1^\star &= \tfrac{(-1)^l}{2}\Big(\ \sqrt{z_1^\star}+\sqrt{z_2^\star}+\sqrt{z_3^\star}\Big)-\tfrac{c_3}{4c_4}, \\
\lambda_2^\star &= \tfrac{(-1)^l}{2}\Big(\ \sqrt{z_1^\star}-\sqrt{z_2^\star}-\sqrt{z_3^\star}\Big)-\tfrac{c_3}{4c_4}, \\
\lambda_3^\star &= \tfrac{(-1)^l}{2}\Big(-\sqrt{z_1^\star}+\sqrt{z_2^\star}-\sqrt{z_3^\star}\Big)-\tfrac{c_3}{4c_4}, \quad \text{and} \\
\lambda_4^\star &= \tfrac{(-1)^l}{2}\Big(-\sqrt{z_1^\star}-\sqrt{z_2^\star}+\sqrt{z_2^\star}\Big)-\tfrac{c_3}{4c_4},
\end{aligned}\right\} \tag{56}$$

where $l \in \{0,1\}$ must be chosen such that [45]

$$(-1)^l\big(\lambda_1^\star\lambda_2^\star\lambda_3^\star+\lambda_1^\star\lambda_2^\star\lambda_4^\star+\lambda_1^\star\lambda_3^\star\lambda_4^\star+\lambda_2^\star\lambda_3^\star\lambda_4^\star\big)=-q.$$

*4) Analytical computation of the roots of a cubic polynomial:* Consider the monic cubic polynomial with real coefficients given by

$$\chi_3(z) := z^3 + d_2z^2 + d_1z + d_0 = 0 \qquad \text{where} \qquad d_2,\ldots,d_0 \in \mathbb{R}. \tag{57}$$

Its roots $z_1^\star$, $z_2^\star$ and $z_3^\star$ can be computed analytically as follows [46, p. 17]. Note that the cubic polynomial $\chi_3(z)$ has (a) one real and a pair of complex conjugate roots if $\tilde{q}^3+\tilde{r}^2>0$, (b) only real roots but at least two are equal if $\tilde{q}^3+\tilde{r}^2=0$, and (c) only real roots but all are distinct if $\tilde{q}^3+\tilde{r}^2<0$. For the following, define

$$\tilde{q} := \frac{d_1}{3}-\frac{d_2^2}{9} \qquad \text{and} \qquad \tilde{r} := \frac{d_1d_2-3d_0}{6}-\frac{d_2^3}{27}.$$

Then, for

$$s_1 := \sqrt[3]{\tilde{r}+\sqrt{\tilde{q}^3+\tilde{r}^2}} \qquad \text{and} \qquad s_2 := \sqrt[3]{\tilde{r}-\sqrt{\tilde{q}^3+\tilde{r}^2}},$$

the three roots of the cubic polynomial (57) are given by

$$z_1^\star = (s_1+s_2)-\tfrac{d_2}{3} \qquad \text{and} \qquad z_{2,3}^\star = -\tfrac{1}{2}(s_1+s_2)-\tfrac{d_2}{3}\pm \jmath\tfrac{\sqrt{3}}{2}(s_1-s_2). \tag{58}$$

## *B. Computation of the quadrics (implicit expressions) for MTPC, MTPV and MTPF*

To derive the implicit forms presented in Sec. III-B, the first two rows of the gradient (41) must be set to zero, i.e.,

$$\boldsymbol{A}\boldsymbol{i}_\mathrm{s}^k+\boldsymbol{a}+\lambda(\boldsymbol{B}\boldsymbol{i}_\mathrm{s}^k+\boldsymbol{b})=\boldsymbol{0}_2. \tag{59}$$

Rewriting this equation component wise, one may solve and eliminate for the Lagrangian multiplier $\lambda$ as follows

$$\begin{aligned}
&\lambda = -\frac{(a_{11},\,a_{12})\boldsymbol{i}_\mathrm{s}^k+a_1}{(b_{11},\,b_{12})\boldsymbol{i}_\mathrm{s}^k+b_1} = -\frac{(a_{12},\,a_{22})\boldsymbol{i}_\mathrm{s}^k+a_2}{(b_{12},\,b_{22})\boldsymbol{i}_\mathrm{s}^k+b_2} \\
&\implies \big((a_{11},\,a_{12})\boldsymbol{i}_\mathrm{s}^k+a_1\big)\big((b_{12},\,b_{22})\boldsymbol{i}_\mathrm{s}^k+b_2\big)-\big((a_{12},\,a_{22})\boldsymbol{i}_\mathrm{s}^k+a_2\big)\big((b_{11},\,b_{12})\boldsymbol{i}_\mathrm{s}^k+b_1\big)=0.
\end{aligned} \tag{60}$$

Re-arranging leads to the following quadric

$$(\boldsymbol{i}_\mathrm{s}^k)^\top\boldsymbol{M}_\mathrm{X}\boldsymbol{i}_\mathrm{s}^k+2\boldsymbol{m}_\mathrm{X}^\top\boldsymbol{i}_\mathrm{s}^k+\mu_\mathrm{X}$$

where

$$\left.\begin{aligned}
\boldsymbol{M}_\mathrm{X}=\boldsymbol{M}_\mathrm{X}^\top &:= \begin{bmatrix} a_{11}b_{12}-a_{12}b_{11} & \tfrac{1}{2}\big(a_{11}b_{22}+a_{12}b_{12}-a_{22}b_{11}-a_{12}b_{12}\big) \\ \tfrac{1}{2}\big(a_{11}b_{22}+a_{12}b_{12}-a_{22}b_{11}-a_{12}b_{12}\big) & a_{12}b_{22}-a_{22}b_{12} \end{bmatrix} \\
&\overset{(38)}{=} \begin{bmatrix} a_{11}b_{12}-a_{12}b_{11} & \tfrac{1}{2}\big(a_{11}b_{22}-a_{22}b_{11}\big) \\ \tfrac{1}{2}\big(a_{11}b_{22}-a_{22}b_{11}\big) & a_{12}b_{22}-a_{22}b_{12} \end{bmatrix}, \\
\boldsymbol{m}_\mathrm{X} &:= \frac{1}{2}\begin{pmatrix} a_{11}b_2+a_1b_{12}-a_{12}b_1-a_2b_{11} \\ a_{12}b_2+a_1b_{22}-a_{22}b_1-a_2b_{12} \end{pmatrix} \quad \text{and} \\
\mu_\mathrm{X} &:= a_1b_2-a_2b_1.
\end{aligned}\right\} \tag{61}$$

The corresponding matrix $\boldsymbol{M}_\mathrm{X}$, vector $\boldsymbol{m}_\mathrm{X}$ and scalar $\mu_\mathrm{X}$ in (59) with $\mathrm{X} \in \{\mathrm{C},\mathrm{V},\mathrm{F}\}$ are obtained for

- the MTPC hyperbola (23) (i.e. $\mathrm{X}=\mathrm{C}$) by setting $\boldsymbol{A}=\boldsymbol{I}_2$, $\boldsymbol{a}=(0,\,0)^\top$, $\boldsymbol{B}=\boldsymbol{T}$ and $\boldsymbol{b}=\boldsymbol{t}$;
- the MTPV hyperbola (29) (i.e. $\mathrm{X}=\mathrm{V}$) by setting $\boldsymbol{A}=\boldsymbol{V}(\omega_\mathrm{k})$, $\boldsymbol{a}=\boldsymbol{v}(\omega_\mathrm{k})$, $\boldsymbol{B}=\boldsymbol{T}$ and $\boldsymbol{b}=\boldsymbol{t}$; and
- the MTPF hyperbola (33) (i.e. $\mathrm{X}=\mathrm{F}$) by setting $\boldsymbol{A}=\boldsymbol{F}$, $\boldsymbol{a}=\boldsymbol{f}$, $\boldsymbol{B}=\boldsymbol{T}$ and $\boldsymbol{b}=\boldsymbol{t}$, with $\boldsymbol{T}$ and $\boldsymbol{t}$ as in (9), $\boldsymbol{V}(\omega_\mathrm{k})$ and $\boldsymbol{v}(\omega_\mathrm{k})$ as in (15) and $\boldsymbol{F}$ and $\boldsymbol{f}$ as in (21), respectively.

*C. Explicit expressions for current circle, voltage ellipse, and torque, MTPC, MTPV and MTPF hyperbolas*

Consider an arbitrary quadric $Q_A(\boldsymbol{i}_s^k)$ as in (37) where $\boldsymbol{A}$, $\boldsymbol{a}$ (and there respective entries $a_{ij}$ and $a_i$) and $\alpha$ are as in (38). An explicit expression for $Q_A(\boldsymbol{i}_s^k)$ can be obtained by solving $Q_A(\boldsymbol{i}_s^k)$ as in (37) for the quadrature current $i_s^q$. To derive an explicit expression, different cases must be taken into account (like signs or whether certain parameters are zero or not) which makes the use of explicit expressions tedious. Assuming the explicit expression of the quadric $Q_A(\boldsymbol{i}_s^k)$ as in (37) exists, it is given by

$$\mathbb{A}(i_s^d) = -\frac{a_{12} i_s^d + a_2}{a_{22}} \pm \frac{\sqrt{\big(a_{12} i_s^d + a_2\big)^2 - a_{22}\big(a_{11}(i_s^d)^2 + 2a_1 i_s^d + \alpha\big)}}{a_{22}}. \tag{62}$$

Clearly, to have a meaningful expression, the following must hold: $a_{22} \neq 0$ and $\big(a_{12} i_s^d + a_2\big)^2 - a_{22}\big(a_{11}(i_s^d)^2 + 2a_1 i_s^d + \alpha\big) \geq 0$ for all $i_s^d \in \mathbb{R}$ (which might not hold in general; leading to different cases where (62) will hold). To compute the explicit expressions for torque, current circle, voltage ellipse, MTPC, MTPF and MTPV hyperbola, the corresponding matrix $\boldsymbol{A}$, vector $\boldsymbol{a}$ and scalar $\alpha$ must be chosen properly (e.g. for the voltage ellipse (17), choose $\boldsymbol{A} = \boldsymbol{V}(\omega_k)$, $\boldsymbol{a} = \boldsymbol{v}(\omega_k)$ and $\alpha = \nu(\omega_k, \hat{u}_{\max})$ with $\boldsymbol{V}(\omega_k)$, $\boldsymbol{v}(\omega_k)$ and $\nu(\omega_k, \hat{u}_{\max})$ as in (15)).

*D. Computation of the intersection points of two arbitrary quadrics*

To find the intersection point(s) $\boldsymbol{x}^\star$ of two arbitrary quadrics $Q_A(\boldsymbol{x})$ and $Q_B(\boldsymbol{x})$ as in (37), there are several possible algorithms. In this paper, an algorithm is presented which leads again to a problem of 'finding the roots of a fourth-order polynomial' and hence can be solved analytically. For the following, introduce

$$\boldsymbol{D} := \begin{bmatrix} d_{11} & d_{12} \\ d_{12} & d_{22} \end{bmatrix} \in \mathbb{R}^{2\times 2},\ \boldsymbol{d} := \begin{pmatrix} d_1 \\ d_2 \end{pmatrix} \in \mathbb{R}^2,\ \boldsymbol{x}_s \in \mathbb{R}^2,\ \boldsymbol{M} := \begin{bmatrix} m_{11} & m_{12} \\ m_{12} & m_{22} \end{bmatrix} \in \mathbb{R}^{2\times 2}, \boldsymbol{m} := \begin{pmatrix} m_1 \\ m_2 \end{pmatrix} \in \mathbb{R}^2, \text{ and } \mu \in \mathbb{R}$$

and the quadric

$$Q_M(\boldsymbol{x}) := \boldsymbol{x}^\top \boldsymbol{M} \boldsymbol{x} + 2\boldsymbol{m}^\top \boldsymbol{x} + \mu \tag{63}$$

To compute the analytical solutions of the intersection points of the quadrics $Q_A(\boldsymbol{x})$ and $Q_B(\boldsymbol{x})$, three different cases must be considered:

(i) If $\alpha \neq 0$ and $\beta \neq 0$ in (37), then, for $\boldsymbol{D} := \big(\frac{\boldsymbol{A}}{\alpha} - \frac{\boldsymbol{B}}{\beta}\big)$ and $\boldsymbol{d} := \big(\frac{\boldsymbol{a}}{\alpha} - \frac{\boldsymbol{b}}{\beta}\big)$, define the difference quadric of the scaled quadrics as follows

$$Q_D(\boldsymbol{x}) := \boldsymbol{x}^\top \boldsymbol{D} \boldsymbol{x} + 2\boldsymbol{d}^\top \boldsymbol{x} = \boldsymbol{x}^\top \Big(\frac{\boldsymbol{A}}{\alpha} - \frac{\boldsymbol{B}}{\beta}\Big)\boldsymbol{x} + 2\Big(\frac{\boldsymbol{a}}{\alpha} - \frac{\boldsymbol{b}}{\beta}\Big)^\top \boldsymbol{x} = 0, \tag{64}$$

and set $\boldsymbol{M} := \boldsymbol{A}$, $\boldsymbol{m} := \boldsymbol{a}$ and $\mu := \alpha$ (or $\boldsymbol{M} := \boldsymbol{B}$, $\boldsymbol{m} := \boldsymbol{b}$ and $\mu := \beta$ in (63)). Both is possible and does not alter the result.

(ii) If $\alpha = 0$ and $\beta \neq 0$ in (37), then set $\boldsymbol{D} := \boldsymbol{A}$, $\boldsymbol{d} := \boldsymbol{a}$ in (64), and $\boldsymbol{M} := \boldsymbol{B}$, $\boldsymbol{m} := \boldsymbol{b}$ and $\mu := \beta$ in (63); *or*
if $\alpha \neq 0$ and $\beta = 0$ in (37), then set $\boldsymbol{D} := \boldsymbol{B}$, $\boldsymbol{d} := \boldsymbol{b}$ in (64), and $\boldsymbol{M} := \boldsymbol{A}$, $\boldsymbol{m} := \boldsymbol{a}$ and $\mu := \alpha$ in (63).

(iii) If $\alpha = 0$ and $\beta = 0$ and/or $\boldsymbol{a} = \boldsymbol{0}_2$ and $\boldsymbol{b} = \boldsymbol{0}_2$ in (37), the quadrics $Q_A(\boldsymbol{x})$ and $Q_B(\boldsymbol{x})$ must be shifted (translated) by $\boldsymbol{x}_s \in \mathbb{R}^2 \setminus \{\boldsymbol{0}_2\}$ by inserting $\boldsymbol{x} := \boldsymbol{y} + \boldsymbol{x}_s$ into (37), i.e.

$$\left.\begin{aligned} Q_A(\boldsymbol{y} + \boldsymbol{x}_s) &:= \boldsymbol{y}^\top \boldsymbol{A} \boldsymbol{y} + 2\underbrace{(\boldsymbol{x}_s^\top \boldsymbol{A} + \boldsymbol{a}^\top)}_{=:\,\boldsymbol{a}_s^\top}\boldsymbol{y} + \underbrace{\boldsymbol{x}_s^\top \boldsymbol{A} \boldsymbol{x}_s + 2\boldsymbol{a}^\top \boldsymbol{x}_s}_{\alpha_s} \text{ and} \\ Q_B(\boldsymbol{y} + \boldsymbol{x}_s) &:= \boldsymbol{y}^\top \boldsymbol{B} \boldsymbol{y} + 2\underbrace{(\boldsymbol{x}_s^\top \boldsymbol{B} + \boldsymbol{b}^\top)}_{=:\,\boldsymbol{b}_s^\top}\boldsymbol{y} + \underbrace{\boldsymbol{x}_s^\top \boldsymbol{B} \boldsymbol{x}_s + 2\boldsymbol{b}^\top \boldsymbol{x}_s}_{\beta_s}, \end{aligned}\right\} \tag{65}$$

such that the shifted scalars and vectors are *non-zero*, i.e. $\alpha_s \neq 0$, $\beta_s \neq 0$, $\boldsymbol{a}_s \neq \boldsymbol{0}_2$ and $\boldsymbol{b}_s \neq \boldsymbol{0}_2$. Then, case (i) holds true again (but now in $\boldsymbol{y}$) and one sets $\boldsymbol{D} := \big(\frac{\boldsymbol{A}}{\alpha_s} - \frac{\boldsymbol{B}}{\beta_s}\big)$, $\boldsymbol{d} := \big(\frac{\boldsymbol{a}_s}{\alpha_s} - \frac{\boldsymbol{b}_s}{\beta_s}\big)$ in (64), $\boldsymbol{M} := \boldsymbol{A}$, $\boldsymbol{m} := \boldsymbol{a}_s$ and $\mu := \alpha_s$ (or $\boldsymbol{M} := \boldsymbol{B}$, $\boldsymbol{m} := \boldsymbol{b}_s$ and $\mu := \beta_s$) in (63). Finally, to obtain the intersections points $\boldsymbol{x}^\star$, the solution $\boldsymbol{y}^\star$ must be translated again, i.e. $\boldsymbol{x}^\star = \boldsymbol{y}^\star + \boldsymbol{x}_s$.

Note that, for all three cases (i), (ii) and (iii)[7], $\boldsymbol{D} = \boldsymbol{D}^\top$ and one may rewrite (64) as follows

$$Q_D(\boldsymbol{x}) = \boldsymbol{x}^\top \boldsymbol{D} \boldsymbol{x} + 2\boldsymbol{d}^\top \boldsymbol{x} = \boldsymbol{x}^\top \big(\underbrace{\boldsymbol{D}\boldsymbol{x} + 2\boldsymbol{d}}_{\overset{!}{=}\gamma \boldsymbol{J}\boldsymbol{x}}\big) = 0. \tag{66}$$

Since the vectors $\boldsymbol{J}\boldsymbol{x}$ (or $\boldsymbol{J}^\top \boldsymbol{x}$) and $\boldsymbol{x}$ are perpendicular to each other, the following holds $\gamma(\boldsymbol{J}\boldsymbol{x})^\top \boldsymbol{x} = \gamma \boldsymbol{x}^\top \boldsymbol{J}^\top \boldsymbol{x} = 0 = \gamma \boldsymbol{x}^\top \boldsymbol{J} \boldsymbol{x}$ for all $\gamma \in \mathbb{R} \setminus \{0\}$ and, so, (66) is clearly satisfied for $\boldsymbol{D}\boldsymbol{x} + 2\boldsymbol{d} \overset{!}{=} \gamma \boldsymbol{J}\boldsymbol{x}$ (the factor $\gamma$ is necessary to allow for scaled versions of the vector $\boldsymbol{J}\boldsymbol{x}$; such that different lengths are admissible). Hence, one obtains

$$[\boldsymbol{D}-\gamma\boldsymbol{J}]\boldsymbol{x}+2\boldsymbol{d}=\boldsymbol{0}_2 \implies \boxed{\boldsymbol{x}(\gamma)=-2[\boldsymbol{D}-\gamma\boldsymbol{J}]^{-1}\boldsymbol{d}} \text{ where } [\boldsymbol{D}-\gamma\boldsymbol{J}]^{-1}=\frac{1}{\gamma^2+d_{11}d_{22}-d_{12}^2}\begin{bmatrix} d_{22}, & -d_{12}-\gamma \\ \gamma-d_{12}, & d_{11}\end{bmatrix}.$$

[7] For case (iii), substitute $\boldsymbol{y}$ for $\boldsymbol{x}$.

Inserting $\boldsymbol{x}(\gamma)$ as above into the quadric $Q_M(\boldsymbol{x})$ as in (63) gives a fourth-order polynomial in $\gamma$, i.e.

$$\left.\begin{array}{rcl} 4\boldsymbol{d}^\top([\boldsymbol{D}-\gamma\boldsymbol{J}]^{-1})^\top\boldsymbol{M}[\boldsymbol{D}-\gamma\boldsymbol{J}]^{-1}\boldsymbol{d}-4\boldsymbol{m}^\top[\boldsymbol{D}-\gamma\boldsymbol{J}]^{-1}\boldsymbol{d}+\mu & = & 0 \quad | \quad \cdot\left(\det[\boldsymbol{D}-\gamma\boldsymbol{J}]\right)^2 \\ \implies \quad \chi_4(\gamma):=\xi_4\gamma^4+\xi_3\gamma^3+\xi_2\gamma^2+\xi_1\gamma+\xi_0 & = & 0 \end{array}\right\} \tag{67}$$

with coefficients

$$\left.\begin{array}{rcl} \xi_4 & := & \mu \\ \xi_3 & := & 4\left(m_1\,d_2-m_2\,d_1\right) \\ \xi_2 & := & 4\,m_{22}\,d_1^2-8\,m_{12}\,d_1\,d_2+4\,m_2\,d_1\,d_{12}-4\,m_1\,d_{22}\,d_1+4\,m_{11}\,d_2^2+4\,m_1\,d_2\,d_{12}-4\,m_2\,d_{11}\,d_2 \\ & & -2\,\mu\,d_{12}^2+2\,\mu\,d_{11}\,d_{22} \\ \xi_1 & := & 4\,m_2\,d_1\,d_{12}^2-4\,m_1\,d_2\,d_{12}^2+8\,m_{11}\,d_2^2\,d_{12}-8\,m_{12}\,d_2^2\,d_{11}+8\,m_{12}\,d_1^2\,d_{22}-8\,m_{22}\,d_1^2\,d_{12} \\ & & +4\,m_1\,d_2\,d_{11}\,d_{22}-4\,m_2\,d_1\,d_{11}\,d_{22}-8\,m_{11}\,d_1\,d_2\,d_{22}+8\,m_{22}\,d_1\,d_2\,d_{11} \\ \xi_0 & := & 4\,m_{22}\,d_1^2\,d_{12}^2-8\,m_{12}\,d_1^2\,d_{12}\,d_{22}+4\,m_{11}\,d_1^2\,d_{22}^2-8\,m_{22}\,d_1\,d_2\,d_{11}\,d_{12}+8\,m_{12}\,d_1\,d_2\,d_{11}\,d_{22} \\ & & +8\,m_{12}\,d_1\,d_2\,d_{12}^2-8\,m_{11}\,d_1\,d_2\,d_{12}\,d_{22}+4\,m_2\,d_1\,d_{11}\,d_{12}\,d_{22}-4\,m_1\,d_1\,d_{11}\,d_{22}^2-4\,m_2\,d_1\,d_{12}^3 \\ & & +4\,m_1\,d_1\,d_{12}^2\,d_{22}+4\,m_{22}\,d_2^2\,d_{11}^2-8\,m_{12}\,d_2^2\,d_{11}\,d_{12}+4\,m_{11}\,d_2^2\,d_{12}^2-4\,m_2\,d_2\,d_{11}^2\,d_{22} \\ & & +4\,m_2\,d_2\,d_{11}\,d_{12}^2+4\,m_1\,d_2\,d_{11}\,d_{12}\,d_{22}-4\,m_1\,d_2\,d_{12}^3+\mu\,d_{11}^2\,d_{22}^2-2\,\mu\,d_{11}\,d_{12}^2\,d_{22}+\mu\,d_{12}^4. \end{array}\right\} \tag{68}$$

The *real* root $\gamma^\star$ of the four roots $\gamma_1^\star$, $\gamma_2^\star$, $\gamma_3^\star$, $\gamma_4^\star$ of the fourth-order polynomial $\chi_4(\gamma)$ as in (67) gives the desired intersection point in the quadrant of interest, i.e.,

$$\text{for cases (i) and (ii): } \boxed{\boldsymbol{x}^\star(\gamma^\star)=-2[\boldsymbol{D}-\gamma^\star\boldsymbol{J}]^{-1}\boldsymbol{d}} \text{ and for case (iii): } \boxed{\boldsymbol{x}^\star(\gamma^\star)=-2[\boldsymbol{D}-\gamma^\star\boldsymbol{J}]^{-1}\boldsymbol{d}+\boldsymbol{x}_\mathrm{s}}. \tag{69}$$

Note that there exist one, two, three or four real roots $\gamma_1^\star$, $\gamma_2^\star$, $\gamma_3^\star$, $\gamma_4^\star$ if the quadrics $Q_A(\boldsymbol{x})$ and $Q_B(\boldsymbol{x})$ do intersect; if the quadrics do not intersect there is no real root.